\documentclass[aps,prd,twocolumn,preprintnumbers,amsmath,amssymb,nofootinbib,floatfix]{revtex4-2}

\usepackage{graphicx}
\usepackage{amsfonts}
\usepackage{booktabs}
\usepackage{multirow}
\usepackage{slashed}
\usepackage{braket}
\usepackage{bm}
\usepackage{xcolor}
\usepackage[colorlinks=true,citecolor=blue,linkcolor=red,urlcolor=blue]{hyperref}

\allowdisplaybreaks[4]

\begin{document}

\title{Probing the two-pole structure of the $D_0^*(2300)$ with an $SU(3)$ flavor filter}

\author{Ying-Bo He$^{1}$}
\email{heyingbo@tju.edu.cn}
\author{Liu-Lin Wang$^{1}$}
\email{liulinwang@tju.edu.cn}
\author{Xiao-Hai Liu$^{1}$}
\email{xiaohai.liu@tju.edu.cn}
\author{Qiang Zhao$^{2,3,4}$}
\email{zhaoq@ihep.ac.cn}

\affiliation{ $^{1}$Center for Joint Quantum Studies and Department of Physics, School of Science, Tianjin University, Tianjin 300350, China\\
$^{2}$ Institute of High Energy Physics and Theoretical Physics Center for Science Facilities,
Chinese Academy of Sciences, Beijing 100049, China\\
$^{3}$ University of Chinese Academy of Sciences, Beijing 100049, China\\
$^{4}$ Center for High Energy Physics, Henan Academy of Sciences, Zhengzhou 450046, China
}

\date{\today}

\begin{abstract}
We propose an $SU(3)$ flavor-filtering strategy to resolve the two-pole structure associated with the $D_0^*(2300)$ and to probe its flavor-sextet component directly in invariant-mass distributions. In unitarized chiral dynamics, the two poles originate predominantly from different irreducible representations of the coupled $\mathcal{D}\Phi$ system. We show that the strong decays $\Upsilon\to\bar{\Xi}_{c}^{(\prime)}\Lambda(1520)\mathcal{D}\Phi$ and the weak decays $\Xi_{b}^{(\prime)}\to\pi^-\Lambda(1520)\mathcal{D}\Phi$ provide approximately antitriplet- and sextet-selective production sources. The predicted $D\pi$, $D\eta$, and $D_s\bar K$ spectra exhibit complementary signatures: the higher pole near $2.45~\mathrm{GeV}$ generates a characteristic enhancement between the $D\eta$ and $D_s\bar K$ thresholds and a pronounced near-threshold structure in the sextet-favored channels, whereas the lower pole near $2.1~\mathrm{GeV}$ dominates the antitriplet-favored channels. The contrast between these line shapes directly reflects the flavor representation selected by the corresponding cascade-baryon tag. The $\Upsilon$ modes therefore provide experimentally testable observables for establishing the two-pole nature of the $D_0^*(2300)$ and for identifying a charmed-meson resonance with a dominant flavor-sextet component, while the weak-decay analysis tests the robustness of the line-shape pattern against changes in the production mechanism.
\end{abstract}

\maketitle

\section{Introduction}

The discovery of numerous hadron candidates that do not fit naturally into conventional $q\bar q$ or $qqq$ spectroscopy has made exotic-hadron studies a major arena for exploring nonperturbative QCD. A recurring lesson is that, particularly near $S$-wave thresholds, hadron--hadron interactions and coupled-channel effects may generate resonance poles or sizable molecular components; consequently, pole positions and channel couplings can be more informative than peak positions alone~\cite{Meng:2022ozq,Mai:2022eur}. This insight is relevant not only to manifestly exotic candidates but also to states with conventional flavor quantum numbers whose masses and line shapes are strongly reshaped by nearby thresholds. The lowest positive-parity open-charm mesons provide a particularly useful testing ground for this interplay between quark-model excitations and hadronic dynamics~\cite{Godfrey:1985xj,DiPierro:2001uu}. The issue is especially acute for the broad nonstrange scalar $D_0^*(2300)$, first isolated by Belle in hadronic $B$ decays and subsequently observed through the $D\pi$ $S$-wave phase motion by LHCb~\cite{Belle:2003nnu,LHCb:2016lxy,ParticleDataGroup:2024cfk}. The central question addressed here is whether the experimental strength conventionally assigned to the $D_0^*(2300)$ is generated by a single resonance or by two nearby poles with different flavor origins.

The narrow charm-strange states $D_{s0}^*(2317)$ and $D_{s1}(2460)$ provide essential context for this question. They were discovered in the isospin-violating $D_s^+\pi^0$ and $D_s^{*+}\pi^0$ channels, respectively, and were confirmed in exclusive $B$ decays~\cite{BaBar:2003oey,CLEO:2003ggt,Belle:2003drx}. The $D_{s0}^*(2317)$ lies about $45~\mathrm{MeV}$ below the $DK$ threshold, and the $D_{s1}(2460)$ lies by a similar amount below the $D^*K$ threshold. Their dominant isospin-conserving $S$-wave decay modes are therefore kinematically closed, naturally accounting for their narrow widths. Their low masses, together with the near degeneracy of the $D_{s0}^*(2317)$ and the broad nonstrange $D_0^*(2300)$ structure, are difficult to reconcile with simple pre-discovery quark-model expectations~\cite{DiPierro:2001uu,Godfrey:2003kg,Godfrey:2005wwa}.

These observations motivated descriptions of the $D_{s0}^*(2317)$ and $D_{s1}(2460)$ as compact chiral partners, $DK$ and $D^*K$ molecules, or coupled-channel states containing both compact and meson--meson components~\cite{Bardeen:2003kt,Barnes:2003dj,Kolomeitsev:2003ac,Guo:2006fu,Guo:2006rp,Cleven:2010aw,Cleven:2014oka}. Lattice-QCD calculations with $D^{(*)}K$ operators find the corresponding subthreshold poles and are consistent with substantial $D^{(*)}K$ components~\cite{Liu:2012zya,Mohler:2013rwa,Lang:2014yfa,MartinezTorres:2014son,Bali:2017pdv}, while Hamiltonian effective-field-theory studies favor mixed core--continuum wave functions~\cite{Yang:2021tvc}. For the present work, the crucial point is that these charm-strange states and the nonstrange positive-parity states emerge as related consequences of the same heavy-meson--Goldstone-boson dynamics. Heavy-quark spin symmetry extends this connection from the scalar sector to the axial-vector sector: lattice and unitarized analyses identify the lower $D_1(2430)$ pole as the nonstrange partner of the $D_{s1}(2460)$, while a second, higher axial pole is related to the scalar sextet pole; femtoscopic correlations have been proposed to test this axial two-pole pattern~\cite{Mohler:2012na,Guo:2006rp,Albaladejo:2016lbb,TorresRincon:2023sdf}. Thus, the $D_{s0}^*(2317)$ and $D_{s1}(2460)$ are closely related benchmarks for the dynamical mechanism studied below, whereas the scalar $D_0^*(2300)$ remains our primary target.

In unitarized chiral perturbation theory, the $I=1/2$, $J^P=0^+$ coupled-channel amplitudes for $D\pi$, $D\eta$, and $D_s\bar K$ contain two poles rather than a single Breit--Wigner resonance~\cite{Albaladejo:2016lbb,Meng:2022ozq,Asokan:2022usm,Meissner:2020khl}. The lower pole, whose real part is close to $2.1~\mathrm{GeV}$, is continuously connected to the attractive flavor-antitriplet interaction in the $SU(3)$ limit and is the nonstrange partner of the $D_{s0}^*(2317)$. The higher and broader pole, located near the $D\eta$ and $D_s\bar K$ thresholds, originates predominantly from the flavor-sextet interaction. This difference is physically significant: a conventional $c\bar q$ meson belongs only to a flavor antitriplet, whereas a sextet arises naturally from the coupled system of a charmed meson and a light pseudoscalar meson. Establishing the two poles would therefore reveal not only the analytic structure behind the broad $D_0^*(2300)$ signal but also a flavor component beyond the conventional quark--antiquark assignment. Since nearby thresholds strongly deform the spectrum, this question must ultimately be addressed through pole and coupled-channel analyses rather than a single Breit--Wigner fit~\cite{Mai:2022eur}.

Lattice-QCD calculations and phenomenological analyses increasingly support this two-pole interpretation~\cite{Moir:2016srx,Du:2017zvv,Gayer:2021xzv,Yan:2024gkx,Yeo:2024chk,Zhuang:2026lta,Luo:2026qen}. Elastic $D\pi$ studies constrain the lower pole and its light-quark-mass dependence, while coupled-channel and $SU(3)$-symmetric analyses reveal the additional structure associated with the higher thresholds. Interest in the sextet component has also been sharpened by the possible exotic $T_{c\bar s}$ signal in the $D_s\pi$ spectrum and by analyses of open-charm flavor multiplets~\cite{LHCb:2024iuo,Maiani:2024quj,Guo:2025vjw}. Semileptonic $B$ decays offer a complementary way to constrain the lower pole through phase information~\cite{Du:2025beb}. Nevertheless, the two contributions generally overlap in physical invariant-mass distributions, and a practical reaction mechanism that changes their relative weights is still required.

Reaction-dependent line shapes are a characteristic manifestation of coupled-channel two-pole dynamics. Related structures have been discussed for the $K_1(1270)$~\cite{Roca:2005nm,Geng:2006yb}, the $D_1(2430)$~\cite{Kolomeitsev:2003ac,Guo:2006fu,Guo:2006rp,Albaladejo:2016lbb,TorresRincon:2023sdf}, the $\Xi(1820)$~\cite{Molina:2023uko}, and especially the $\Lambda(1405)$~\cite{Hyodo:2011ur,Xie:2023two}. In the latter case, attractive singlet and octet interactions in the $\mathbf 8\otimes\mathbf 8$ meson--baryon system generate poles with different couplings to the physical channels, leading naturally to production-dependent $\pi\Sigma$ spectra~\cite{Oller:2000fj,Jido:2003cb}. NLO studies have followed their evolution away from the $SU(3)$ limit~\cite{Guo:2023mku}, and coupled-channel lattice calculations have resolved the two-pole pattern and, more recently, its singlet--octet content at an $SU(3)$-symmetric point~\cite{Bulava:2023uma,SuarezSucunza:2026lambda}.

Ref.~\cite{He:2024uau} converted this reaction dependence into an $SU(3)$-representation filter for the $\Lambda(1405)$. Because a heavy quarkonium source is a light-flavor singlet, the accompanying antibaryon in $Y\to\bar\Lambda^{(*)}\Sigma\pi$ fixes the representation of the initially produced meson--baryon pair: the $\bar\Lambda$ and predominantly singlet $\bar\Lambda(1520)$ tags generate complementary octet- and singlet-dominated $\pi\Sigma$ spectra. The distinction survives physical $SU(3)$ breaking, although higher-order pole trajectories caution against identifying a physical pole too rigidly with one symmetric-limit representation~\cite{Guo:2023mku,He:2024uau}. This result provides the direct methodological starting point for the present study.

Here we apply the same representation-tagging idea to the scalar charmed-meson sector, for which $\bar{\mathbf 3}\otimes\mathbf 8=\bar{\mathbf{15}}\oplus\mathbf 6\oplus\bar{\mathbf 3}$. We consider the strong decays $\Upsilon\to\bar\Xi_c^{(\prime)}\Lambda(1520)\mathcal{D}\Phi$ and the weak decays $\Xi_b^{(\prime)}\to\pi^-\Lambda(1520)\mathcal{D}\Phi$. In the strong decays, the singlet character of the $\Upsilon$ and, to a good approximation, the $\Lambda(1520)$ allows the accompanying $\bar\Xi_c$ or $\bar\Xi_c'$ to tag an antitriplet- or sextet-dominated $\mathcal{D}\Phi$ source. In the weak decays, the light-diquark representation of the initial $\Xi_b$ or $\Xi_b'$ provides the analogous tag because it is preserved at leading order in the heavy-quark transition. We calculate the $D\pi$, $D\eta$, and $D_s\bar K$ invariant-mass distributions and follow their evolution from the $SU(3)$-symmetric limit to the physical point, with the aim of separating the lower antitriplet and higher sextet poles. Section~\ref{sec:formalism} introduces the coupled-channel amplitude and the strong and weak production mechanisms, Sec.~\ref{sec:results} presents the resulting flavor-filtered spectra, and the main conclusions are summarized in Sec.~\ref{sec:summary}.

\section{Formalism}\label{sec:formalism}

\subsection{Two-pole structure of the $D_0^*(2300)$}

Throughout this work, $\mathcal{D}\Phi$ is used as a collective label for a two-meson channel composed of a charmed pseudoscalar meson $\mathcal{D}\in\{D,D_s\}$ and a light pseudoscalar Nambu--Goldstone boson $\Phi\in\{\pi,K,\bar K,\eta\}$. In expressions with explicit $SU(3)$ flavor indices, $\mathcal{D}$ and $\Phi$ denote, respectively, the charmed-meson antitriplet column and the light-pseudoscalar octet matrix introduced below. When written without flavor indices in reaction labels, invariant masses, or prose, $\mathcal{D}\Phi$ is instead shorthand for the coupled two-meson system rather than a literal product of these multiplet fields.

The $S$-wave interaction of a charmed pseudoscalar meson with the light Nambu--Goldstone octet is described by the on-shell unitarized chiral amplitude
\begin{equation}
T^{-1}(s)=V^{-1}(s)-G(s),
\label{eq:unitarized}
\end{equation}
where $s=M_{\mathcal{D}\Phi}^2$, $V(s)$ is the matrix of $S$-wave-projected interaction kernels, and $G(s)$ is the diagonal matrix of two-meson loop functions. Before $S$-wave projection, the interaction kernel consists of the leading-order (LO) Weinberg--Tomozawa term and next-to-leading-order (NLO) contact terms,
\begin{align}
V(s,t,u)=&\frac{1}{F^2}\bigg[\frac{C_{\rm LO}}{4}(s-u)-4C_0h_0+2C_1h_1 \notag\\
&-2C_{24}H_{24}(s,t,u)+2C_{35}H_{35}(s,t,u)\bigg].
\label{eq:potential}
\end{align}
Here $s$, $t$, and $u$ are the Mandelstam variables, $F$ is the pseudoscalar decay constant, and $h_n$ are NLO low-energy constants. Their numerical values and the channel-dependent coefficients are collected in Appendix~\ref{sec:appendix}. They follow from the LO and NLO chiral Lagrangians~\cite{Guo:2008gp,Guo:2009ct,Liu:2012zya,Altenbuchinger:2013vwa,Yao:2015qia,Du:2016puz}; related global analyses also test the effects of higher-order terms, renormalization prescriptions, and left-hand cuts~\cite{Lutz:2022dcl}. In the sector with strangeness $S=0$ and isospin $I=1/2$, there are three channels $(D\pi,D\eta,D_s\bar K)$.

The group-theoretical origin of the two poles is transparent in the $SU(3)$ limit. Since the charmed-meson multiplet transforms as $\bar{\mathbf 3}$ and the Nambu--Goldstone bosons as $\mathbf 8$,
\begin{equation}
\bar{\mathbf 3}\otimes\mathbf 8
=\bar{\mathbf{15}}\oplus\mathbf 6\oplus\bar{\mathbf 3}.
\label{eq:decomposition}
\end{equation}
At LO, transforming the coefficient matrix to the irreducible-representation basis gives
\begin{equation}
C_{\alpha\beta}^{SU(3)}
=\sum_{i,j}U_{\alpha i}C_{ij}U^\dagger_{j\beta}
=\operatorname{diag}(1,-1,-3),
\label{eq:cg}
\end{equation}
where $U$ is the unitary matrix of the relevant $SU(3)$ Clebsch--Gordan coefficients and the ordering is $\bar{\mathbf{15}}$, $\mathbf 6$, and $\bar{\mathbf 3}$. In the sign convention of Eq.~\eqref{eq:potential}, the negative eigenvalues identify attractive sextet and antitriplet interactions, with the antitriplet being more attractive. Analytically continuing $T(s)$ to the appropriate unphysical Riemann sheets gives the physical-point pole positions
\begin{align}
\sqrt{s_{p,\bar 3}}&\simeq(2105-i\,102)~\mathrm{MeV},
\label{eq:pole-antitriplet}\\
\sqrt{s_{p,6}}&\simeq(2451-i\,134)~\mathrm{MeV}.
\label{eq:pole-sextet}
\end{align}
For brevity, we refer to them below as the lower $D_0^*(2100)$ pole and the higher $D_0^*(2400)$ pole, without implying that they are independent Breit--Wigner states.

To track their evolution under flavor-symmetry breaking, the light- and charmed-meson masses are interpolated according to
\begin{align}
m_i(x)&=m+x\bigl(m_i^{\rm phys}-m\bigr),
\label{eq:su3breaking-light}\\
M_i(x)&=M+x\bigl(M_i^{\rm phys}-M\bigr).
\label{eq:su3breaking-charm}
\end{align}
Here $0\leq x\leq 1$, and $m_i$ and $M_i$ denote the light- and charmed-meson masses, respectively, in channel $i$. We use the common symmetric-point values $m=0.49~\mathrm{GeV}$ and $M=1.95~\mathrm{GeV}$, following Ref.~\cite{Albaladejo:2016lbb}. The point $x=0$ is the $SU(3)$ limit, whereas $x=1$ corresponds to the physical masses. In the symmetric limit, the antitriplet pole is connected to a bound state and the sextet pole to a virtual state. Their trajectories toward the physical point provide a useful diagnostic of the flavor content of the observed line shapes.

\subsection{Production in strong $\Upsilon(nS)$ decays}

\begin{figure}[tb]
\centering
\includegraphics[width=\linewidth]{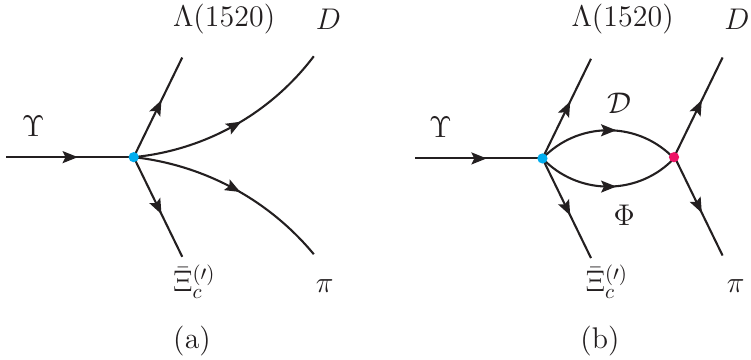}
\caption{Representative mechanisms for the $D\pi$ final state in $\Upsilon\to\bar\Xi_c^{(\prime)}\Lambda(1520)D\pi$: (a) direct production and (b) production of an intermediate $\mathcal{D}\Phi$ pair followed by coupled-channel rescattering $\mathcal{D}\Phi\to D\pi$. Here $\mathcal{D}$ collectively denotes $D$ or $D_s$, while $\Phi$ denotes a light Nambu--Goldstone boson; analogous diagrams apply to the $D\eta$ and $D_s\bar K$ final states.}
\label{fig:strong}
\end{figure}

We first consider the bottomonium four-body decays $\Upsilon(nS)\to\bar\Xi_c^{(\prime)}\Lambda(1520)\mathcal{D}\Phi$. The $\Lambda(1520)$, with quantum numbers $J^P=3/2^-$, is conventionally identified with the predominantly flavor-singlet member of the negative-parity $SU(6)\otimes O(3)$ baryon multiplet~\cite{Klempt:2020bdu,ParticleDataGroup:2024cfk}. Although $SU(3)$ breaking permits mixing with octet configurations, quark-model wave functions retain a large ${}^{2}\mathbf 1[70]$ singlet component~\cite{Schat:2001xr,Melde:2008yr,Klempt:2020bdu}. In addition, the relative $\bar K N$ and $\pi\Sigma$ residue phases extracted in multichannel analyses are consistent with the singlet decay-coupling pattern~\cite{Zhang:2013sva,Kamano:2015hxa,Matveev:2019igl,Sarantsev:2019xxm,Klempt:2020bdu}. In this work, we treat the $\Lambda(1520)$ as a flavor singlet. The vector bottomonium $\Upsilon$ is also a light-flavor singlet. In the ideal flavor basis used in the calculation, $\Xi_c$ and $\Xi_c'$ belong to the $\bar{\mathbf 3}$ and $\mathbf 6$, respectively, and the corresponding antibaryons belong to the $\mathbf 3$ and $\bar{\mathbf 6}$~\cite{Cheng:2006dk}. Flavor conservation therefore correlates $\bar\Xi_c$ production with a $\mathcal{D}\Phi$ source in $\bar{\mathbf 3}$ and $\bar\Xi_c'$ production with a source in $\mathbf 6$. Thus, $\Upsilon\to\bar\Xi_c\Lambda(1520)\mathcal{D}\Phi$ favors the lower antitriplet pole, whereas $\Upsilon\to\bar\Xi_c'\Lambda(1520)\mathcal{D}\Phi$ favors the higher sextet pole. At physical meson masses, final-state rescattering couples the $D\pi$, $D\eta$, and $D_s\bar K$ channels while retaining this representation-selective short-distance source. The production amplitude contains both a direct tree-level term and a final-state-interaction (FSI) contribution, represented by the two diagrams in Fig.~\ref{fig:strong}.

To describe the tree-level short-distance production vertices, we introduce effective Lagrangians constrained by Lorentz invariance, parity, and light-flavor $SU(3)$ symmetry, retaining the minimal derivative dependence required by the external quantum numbers. We first collect the field conventions needed to display the flavor contractions. In this context, the light-pseudoscalar octet field $\Phi$ is represented by
\begin{equation}
\Phi=\frac{1}{\sqrt2}
\begin{pmatrix}
\pi^0+\eta/\sqrt3 & \sqrt2\pi^+ & \sqrt2K^+\\
\sqrt2\pi^- & -\pi^0+\eta/\sqrt3 & \sqrt2K^0\\
\sqrt2K^- & \sqrt2\bar K^0 & -2\eta/\sqrt3
\end{pmatrix},
\label{eq:phi}
\end{equation}
and the charmed-meson antitriplet column $\mathcal D$ is written as~\cite{Guo:2009ct}
\begin{equation}
	\mathcal D=
	\begin{pmatrix}D^{0}\\ D^{+}\\ D_s^{+}\end{pmatrix}.
	\label{eq:charmedmesonmultiplet}
\end{equation}
The charmed-baryon multiplets are~\cite{Cheng:2006dk}
\begin{align}
B_6&=
\begin{pmatrix}
\Sigma_c^{++} & \Sigma_c^+/\sqrt2 & \Xi_c^{\prime+}/\sqrt2\\
\Sigma_c^+/\sqrt2 & \Sigma_c^0 & \Xi_c^{\prime0}/\sqrt2\\
\Xi_c^{\prime+}/\sqrt2 & \Xi_c^{\prime0}/\sqrt2 & \Omega_c^0
\end{pmatrix},
\label{eq:baryonmultiplet-sextet}\\
B_{\bar 3}&=\bigl(\Xi_c^0,-\Xi_c^+,\Lambda_c^+\bigr).
\label{eq:baryonmultiplet-antitriplet}
\end{align}

The physical cascade baryons can contain admixtures of the two light-diquark flavor representations because of $SU(3)$ breaking. Lattice-QCD and QCD-sum-rule analyses indicate, however, that the $\Xi_Q$--$\Xi_Q'$ mixing is small~\cite{Liu:2023feb,Sun:2023noo,Aliev:2010ra}. In the present calculation, this mixing is neglected: $\Xi_Q$ and $\Xi_Q'$ are treated as pure antitriplet and sextet states, respectively, for $Q=c,b$. Treating the $\mathcal{D}\Phi$ pair as an $S$-wave scalar subsystem with $J^P=0^+$, the local interactions can be written as
\begin{equation}
\mathcal L_{6}
=
g_6\,
\varepsilon^{\mu\nu\rho\sigma}
(\partial_\sigma\Upsilon_\nu)\,
\epsilon_{jkl}
\left(
\bar\Lambda_\mu\gamma_5
\overleftrightarrow{\partial}_{\rho}
B_6^{ij}
\right)
\Phi_i{}^{l}\mathcal D^{k}
+\mathrm{H.c.},
\label{eq:Lagstrongsextet}
\end{equation}

\begin{equation}
\mathcal L_{\bar3}
=
g_{\bar3}\,
\varepsilon^{\mu\nu\rho\sigma}
(\partial_\sigma\Upsilon_\nu)
\left(
\bar\Lambda_\mu\gamma_5
\overleftrightarrow{\partial}_{\rho}
B_{\bar3,i}
\right)
\Phi^{i}{}_{j}\mathcal D^{j}
+\mathrm{H.c.},
\label{eq:Lagstrongtriplet}
\end{equation}
Here $\Lambda_\mu$ is the Rarita--Schwinger field for the $\Lambda(1520)$, $g_6$ and $g_{\bar3}$ are coupling constants, and $A\overleftrightarrow{\partial}_{\rho}B\equiv A(\partial_\rho B)-(\partial_\rho A)B$. The tensors $\varepsilon^{\mu\nu\rho\sigma}$ and $\epsilon_{jkl}$ are the Levi--Civita tensors in Lorentz and flavor space, respectively. After factoring out the overall coupling associated with representation $R$ and the common
Lorentz--spin--momentum structure, the relative strengths and
phases of the tree-level amplitudes in the ordered isospin-$1/2$
basis $(D\pi,D\eta,D_s\bar K)$ are encoded in the dimensionless
effective couplings $h_i^{(R)}$. These couplings are fixed by the
$SU(3)$ flavor contractions in Eqs.~\eqref{eq:Lagstrongsextet}
and \eqref{eq:Lagstrongtriplet}, which give
\begin{align}
 h^{(6)}&\equiv \left(\begin{array}{c}
	h_{D\pi}^{(6)} \\[7pt]
	h_{D\eta}^{(6)}\\[7pt]
	h_{D_s\bar{K}}^{(6)}
\end{array}\right)=\left(\begin{array}{c}
	\sqrt{\frac{3}{8}} \\
	-\sqrt{\frac{3}{8}}\\
	-\frac{1}{2}
\end{array}\right),
\label{eq:sourcevector-sextet}\\[2mm]
 h^{(\bar{3})}&\equiv \left(\begin{array}{c}
	h_{D\pi}^{({\bar{3}})} \\[7pt]
	h_{D\eta}^{({\bar{3}})}\\[7pt]
	h_{D_s\bar{K}}^{({\bar{3}})}
\end{array}\right)=\left(\begin{array}{c}
	-\frac{3}{4} \\[4pt]
	-\frac{1}{4}\\[4pt]
	-\sqrt{\frac{3}{8}}
\end{array}\right).
\label{eq:sourcevector-antitriplet}
\end{align}

The $SU(3)$-symmetric tree-level amplitudes $\mathcal M_{i,(6)}^{\mathrm{tree}}$ and $\mathcal M_{i,(\bar 3)}^{\mathrm{tree}}$ follow from Eqs.~\eqref{eq:Lagstrongsextet}
and \eqref{eq:Lagstrongtriplet}. Including the $\mathcal{D}\Phi$ FSI, the amplitude for producing channel $i$ becomes
\begin{equation}
\begin{aligned}
\mathcal M_{i,R}^{\mathrm{tot}}
&=\mathcal M_{i,R}^{\mathrm{tree}}
+\sum_j\mathcal M_{j,R}^{\mathrm{tree}}G_j(s)T_{ji}(s).
\end{aligned}
\label{eq:Adecay}
\end{equation}
Here $s=M_{\mathcal{D}\Phi}^2$ and $i,j\in\{D\pi,D\eta,D_s\bar K\}$,
$G_j$ is the two-meson loop function, and $T_{ji}$ is the unitarized $\mathcal{D}\Phi$ scattering amplitude obtained from Ref.~\cite{Albaladejo:2016lbb}. The relevant numerical parameters are summarized in Appendix~\ref{sec:appendix}.

\subsection{Production in weak $\Xi_b^{(\prime)}$ decays}

\begin{figure}[tb]
\centering
\includegraphics[width=0.80\linewidth]{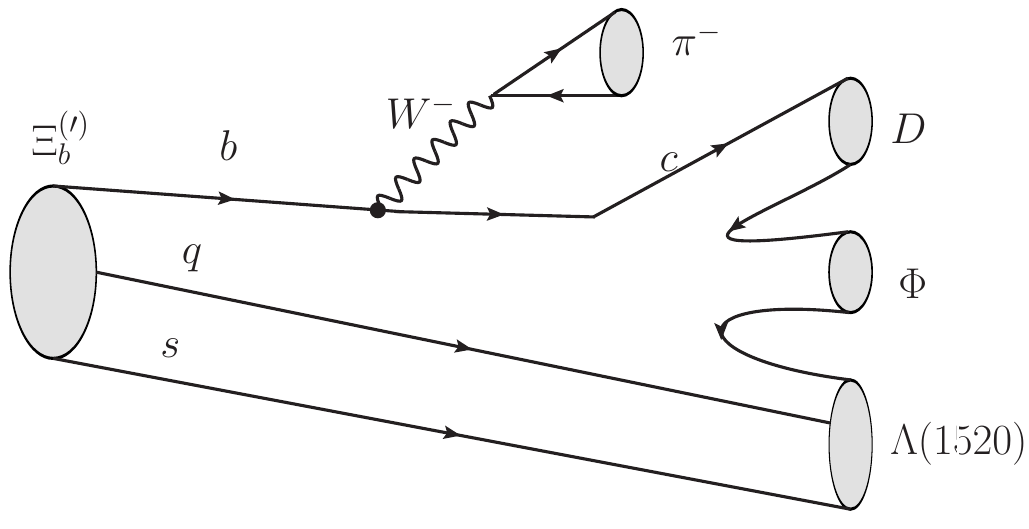}
\caption{Quark-level external-$W$-emission topology for $\Xi_b^{(\prime)}\to\pi^-\Lambda(1520)\mathcal{D}\Phi$. The $b$ quark undergoes $b\to cW^-$ followed by $W^-\to d\bar u$, while the light quarks $q$ and $s$ are spectators. The resulting $cqs$ system forms an off-shell $\Xi_c^{(\prime)}$, which subsequently produces the $\Lambda(1520)\mathcal{D}\Phi$ final state.}
\label{fig:quark}
\end{figure}

\begin{figure}[tb]
\centering
\includegraphics[width=\linewidth]{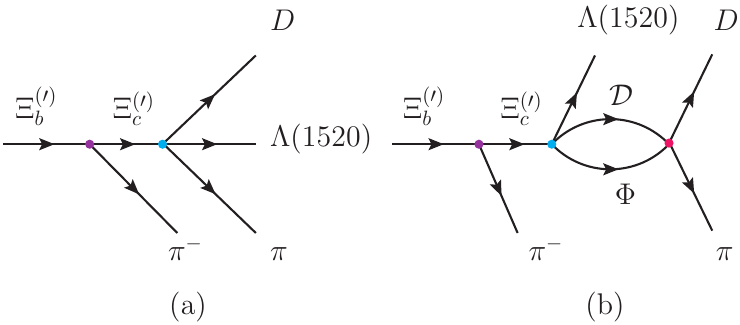}
\caption{Representative mechanisms for the $D\pi$ final state in $\Xi_b^{(\prime)}\to\pi^-\Lambda(1520)D\pi$, mediated by an off-shell $\Xi_c^{(\prime)}$: (a) direct $D\pi$ production and (b) production of an intermediate $\mathcal{D}\Phi$ pair followed by coupled-channel rescattering $\mathcal{D}\Phi\to D\pi$. Analogous diagrams apply to the $D\eta$ and $D_s\bar K$ final states.}
\label{fig:weak}
\end{figure}

We next study $\Xi_b^{(\prime)}\to\pi^-\Lambda(1520)\mathcal{D}\Phi$. At the quark level, the dominant external-emission topology is $b\to cW^-$ followed by $W^-\to d\bar u$, while the two light quarks in $\Xi_b^{(\prime)}$ act as spectators; see Fig.~\ref{fig:quark}. In the heavy-quark limit, the flavor representation of the light diquark is therefore preserved by the weak transition. The resulting intermediate charmed baryon is off shell and subsequently couples to $\Lambda(1520)\mathcal{D}\Phi$. With the ideal flavor assignment specified above, an initial $\Xi_b$ carries an antitriplet light diquark and produces an antitriplet $\mathcal{D}\Phi$ source, whereas an initial $\Xi_b'$ carries a sextet light diquark and produces a sextet source. The direct and rescattering topologies are displayed in Fig.~\ref{fig:weak}.

Within naive factorization, we use the weak current $J^\mu=\bar c\gamma^\mu(1-\gamma_5)b$ and decompose the representation-preserving baryonic matrix element following Ref.~\cite{Ke:2024aux}. In the notation of that reference, the superscripts $s$ and $v$ label scalar (spin-0) and axial-vector (spin-1) spectator-diquark configurations, corresponding here to the antitriplet and sextet flavor bases, respectively. We therefore identify $F^{(\bar3)}\equiv F^s$ and $F^{(6)}\equiv F^v$. For $R\in\{\mathbf6,\bar{\mathbf3}\}$, the pure-representation weak vertex is parameterized as
\begin{align}
&\langle\Xi_c^R(p',\lambda')|J^\mu|\Xi_b^R(p,\lambda)\rangle\notag\\
&\quad=\bar u_{\Xi_c^R}(p',\lambda')
\bigl[\Gamma_{V,R}^\mu(q^2)-\Gamma_{A,R}^\mu(q^2)\gamma_5\bigr]
 u_{\Xi_b^R}(p,\lambda),
\label{eq:weakcurrent}
\end{align}
where $p$ ($p'$) and $\lambda$ ($\lambda'$) denote the momentum and spin projection of the initial (intermediate) baryon, and $q=p-p'=p_\pi$. The vertex functions are
\begin{align}
\Gamma_{V,R}^\mu(q^2)={}&\gamma^\mu f_1^{(R)}(q^2)
+i\sigma^{\mu\nu}\frac{q_\nu}{M_b}f_2^{(R)}(q^2)\notag\\
&+\frac{q^\mu}{M_b}f_3^{(R)}(q^2),
\label{eq:vector-vertex}\\
\Gamma_{A,R}^\mu(q^2)={}&\gamma^\mu g_1^{(R)}(q^2)
+i\sigma^{\mu\nu}\frac{q_\nu}{M_b}g_2^{(R)}(q^2)\notag\\
&+\frac{q^\mu}{M_b}g_3^{(R)}(q^2).
\label{eq:axial-vertex}
\end{align}
Here $M_b$ is the mass of the initial bottom cascade baryon in the corresponding representation. The pion is created by the light-quark current,
\begin{equation}
\langle\pi^-(q)|\bar d\gamma^\mu(1-\gamma_5)u|0\rangle=-if_\pi q^\mu,
\label{eq:pioncurrent}
\end{equation}
where $f_\pi$ is the pion decay constant. Upon contraction with the pion current, the contributions proportional to $f_3^{(R)}$ and $g_3^{(R)}$ carry the factor $q^2/M_b=m_\pi^2/M_b$, whereas the leading terms are set by a much larger baryonic momentum scale; we therefore neglect $f_3^{(R)}$ and $g_3^{(R)}$. The remaining form factors $F^{(R)}\in\{f_1^{(R)},f_2^{(R)},g_1^{(R)},g_2^{(R)}\}$ are parameterized as
\begin{equation}
F^{(R)}(q^2)=F^{(R)}(0)\left[1+a\frac{q^2}{M_b^2}
+b\left(\frac{q^2}{M_b^2}\right)^2\right],
\label{eq:ffparam}
\end{equation}
The form-factor-specific values of $F^{(R)}(0)$, $a$, and $b$ from Ref.~\cite{Ke:2024aux} are listed in Table~\ref{tab:formfactors}. The numerical weak-production amplitude is evaluated at $q^2=m_\pi^2$. Consistently with the pure-flavor treatment stated above, transitions between different representations are set to zero at this order:
\begin{equation}
\langle\Xi_c^R|J^\mu|\Xi_b^{R'}\rangle=0,
\qquad R\ne R'.
\label{eq:weakflavorselection}
\end{equation}
The intermediate $\Xi_c^{(\prime)}$ is far off shell over the kinematic region of interest. We use the standard spin-$1/2$ baryon propagator and neglect its small physical width. Its subsequent strong coupling to $\Lambda(1520)\mathcal{D}\Phi$ is described by representation-preserving local vertices. Treating the $S$-wave $\mathcal{D}\Phi$ pair as a scalar subsystem, the leading interactions with the minimal derivative dependence are

\begin{subequations}
\label{eq:loopamp}
\begin{align}
\mathcal L_{6}^{\prime}
={}&
g_6^{\prime}\,\epsilon_{jkl}\,
\bar\Lambda^\mu B_6^{ij}\,
\partial_\mu
\left(\Phi_i{}^{l}\mathcal D^k\right)
+\mathrm{H.c.},
\label{eq:loopamp6}
\\
\mathcal L_{\bar3}^{\prime}
={}&
g_{\bar3}^{\prime}\,
\bar\Lambda^\mu B_{\bar3,i}\,
\partial_\mu
\left(\Phi^{i}{}_{j}\mathcal D^j\right)
+\mathrm{H.c.}
\label{eq:loopampbar3}
\end{align}
\end{subequations}
The $\mathcal{D}\Phi$ FSI is then included according to Eq.~\eqref{eq:Adecay}, with the appropriate weak-production source replacing the strong tree-level source.

For both the strong and weak production mechanisms, the invariant-mass
distribution is calculated from the Lorentz-invariant four-body phase
space,
\begin{equation}
d\Gamma
=
\frac{1}{2M_A}\,
\overline{|\mathcal M|^2}\,
d\Phi_4,
\label{eq:phase-space}
\end{equation}
where $M_A$ is the mass of the parent hadron, and the bar denotes an average over its spin or polarization and a sum over final-state spins. The four-body phase-space element is
\begin{equation}
d\Phi_4
=
(2\pi)^4
\delta^{(4)}
\left(P-\sum_{i=1}^{4}p_i\right)
\prod_{i=1}^{4}
\frac{d^3\bm p_i}{(2\pi)^3 2E_i},
\label{eq:fourbodyphase}
\end{equation}
where $P$ is the four-momentum of the parent hadron and $p_i=(E_i,\bm p_i)$ is the four-momentum of final-state particle $i$. The phase-space integral is evaluated by weighted Monte Carlo sampling, and $d\Gamma/dM_{\mathcal{D}\Phi}$ is obtained by binning the generated events in $M_{\mathcal{D}\Phi}$. Physical hadron masses are used, with isospin-averaged masses in the coupled-channel amplitude. The $\Lambda(1520)$ is treated in the narrow-width approximation; its subsequent decay does not affect the $\mathcal{D}\Phi$ flavor-filter argument.

\begin{table*}[t]
\centering
\caption{Form-factor parameters in the scalar-diquark ($s$) and axial-vector-diquark ($v$) notation of Ref.~\cite{Ke:2024aux}.}
\label{tab:formfactors}
\begin{tabular}{lrrrrrrrr}
\toprule
& $f_1^s$ & $f_2^s$ & $g_1^s$ & $g_2^s$ & $f_1^v$ & $f_2^v$ & $g_1^v$ & $g_2^v$\\
\midrule
$F(0)$ & 0.467 & -0.185 & 0.448 & -0.052 & 0.471 & 0.378 & -0.149 & -0.006\\
$a$    & 2.19  & 2.70   & 2.09  & 3.32   & 2.57  & 2.56  & 2.07   & 2.95\\
$b$    & 2.09  & 3.01   & 1.92  & 4.54   & 2.40  & 2.81  & 2.00   & 2.98\\
\bottomrule
\end{tabular}
\end{table*}

\section{Results and discussion}\label{sec:results}

Because the short-distance production couplings $g_R^{(\prime)}$ are unknown, neither the absolute branching fractions nor the relative normalizations of different reactions can be predicted. We therefore normalize each spectrum independently and focus on its energy-dependent line shape. In particular, peak heights should not be compared between different panels; the relevant observables are the locations and shapes of the enhancements and cusps, together with the redistribution of spectral strength when the flavor tag is changed.

\begin{figure*}[t]
\begin{minipage}[t]{0.48\textwidth}
\centering
\includegraphics[width=0.88\linewidth]{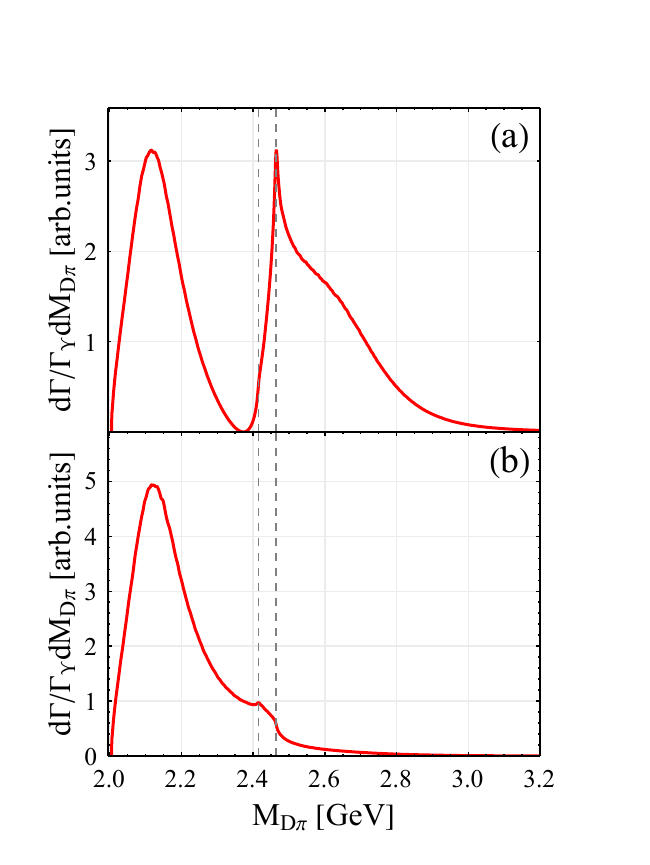}
\caption{Normalized $D\pi$ invariant-mass distributions for (a) $\Upsilon\to\bar\Xi_c'\Lambda(1520)D\pi$ and (b) $\Upsilon\to\bar\Xi_c\Lambda(1520)D\pi$. The vertical dashed lines indicate the $D\eta$ and $D_s\bar K$ thresholds.}
\label{fig:strongdpi}
\end{minipage}\hfill
\begin{minipage}[t]{0.48\textwidth}
\centering
\includegraphics[width=0.88\linewidth]{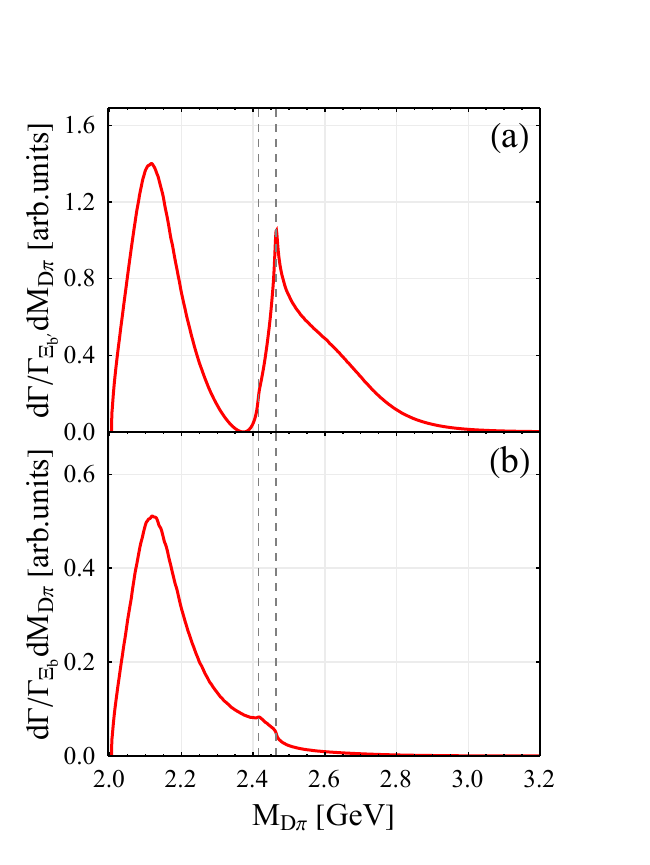}
\caption{Normalized $D\pi$ invariant-mass distributions for (a) $\Xi_b'\to\pi^-\Lambda(1520)D\pi$ and (b) $\Xi_b\to\pi^-\Lambda(1520)D\pi$. The vertical dashed lines indicate the $D\eta$ and $D_s\bar K$ thresholds.}
\label{fig:weakdpi}
\end{minipage}
\end{figure*}

Figure~\ref{fig:strongdpi} displays the $D\pi$ spectra in the strong $\Upsilon$ decays. We first focus on the sextet-favored channel in Fig.~\ref{fig:strongdpi}(a), because it provides the clearest test of the proposed flavor filter. Its salient feature is the higher-mass structure associated with the higher, predominantly sextet pole: the distribution rises steeply above the $D\eta$ threshold, is strongly enhanced in the interval bounded by the $D\eta$ and $D_s\bar K$ thresholds, and develops a pronounced cusp at the opening of the $D_s\bar K$ channel, followed by a broad high-mass shoulder. Since the pole is broad and lies close to two thresholds, it is not expected to generate a symmetric Breit--Wigner peak. The prominence of this threshold-distorted structure in the $\bar\Xi_c'$-tagged reaction is precisely the behavior expected when the short-distance source projects predominantly onto the flavor sextet.

The same panel also exhibits a broad low-mass enhancement around $2.1~\mathrm{GeV}$. Its presence does not imply that the source has lost its flavor selectivity. In the exact $SU(3)$ limit, the sextet source $h^{(6)}$ excites only the sextet amplitude and is orthogonal to the lower antitriplet pole. At the physical meson masses, however, the unequal masses and thresholds make both the loop functions and the interaction kernels channel dependent, thereby mixing the irreducible representations in the coupled-channel amplitude. The nominally sextet source can consequently feed the lower pole, whose strong coupling to the open $D\pi$ channel makes this $SU(3)$-breaking leakage clearly visible. Thus, at the physical point the filter is selective rather than exclusive: the higher pole is preferentially enhanced, while the lower pole can still appear through flavor-symmetry breaking.

The complementary behavior is seen in the antitriplet-favored channel in Fig.~\ref{fig:strongdpi}(b). Here the broad lower-pole enhancement dominates, whereas the higher-energy region is strongly reduced and retains only a modest structure associated with the nearby thresholds. Switching the flavor tag from $\bar\Xi_c'$ to $\bar\Xi_c$ therefore changes the relative weights of the two mass regions. This paired comparison, rather than the observation of either enhancement in isolation, constitutes the central flavor-filter signature.

Figure~\ref{fig:weakdpi} provides an independent realization of the same test in weak $\Xi_b^{(\prime)}$ decays. The sextet-tagged $\Xi_b'$ channel in panel (a) again displays the higher-pole structure together with the $SU(3)$-breaking lower-pole contribution, whereas the antitriplet-tagged $\Xi_b$ channel in panel (b) is dominated by the lower pole. The parent-hadron phase space, weak form factors, and off-shell baryon propagator supply different smooth production weights, but the positions of the enhancements and the threshold-cusp pattern remain stable because both classes of reactions contain the same coupled-channel $\mathcal{D}\Phi$ amplitude. The recurrence of the pattern in two distinct production mechanisms shows that the proposed filter does not rely on a finely tuned short-distance vertex.

\begin{figure*}[t]
\begin{minipage}[t]{0.48\textwidth}
\centering
\includegraphics[width=0.88\linewidth]{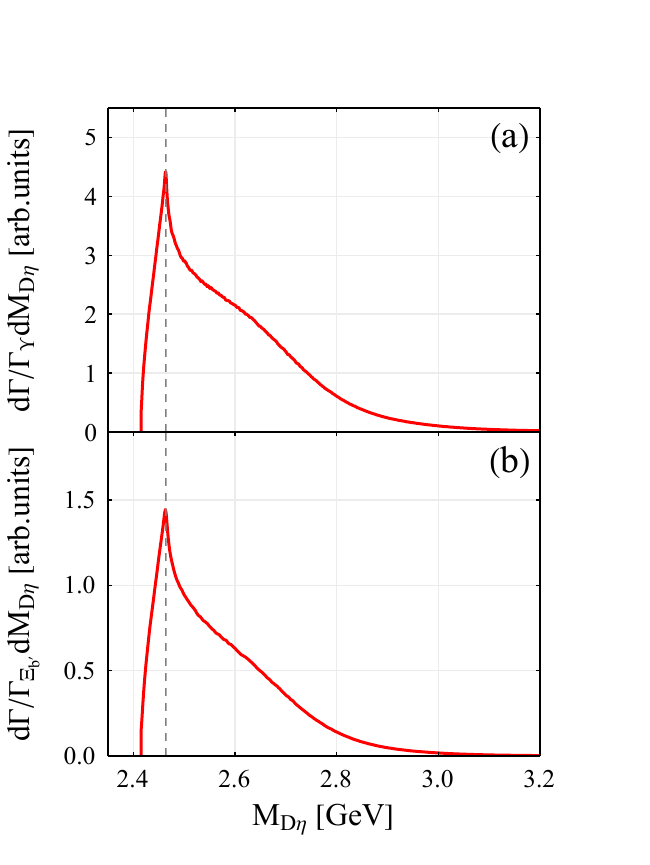}
\caption{Normalized $D\eta$ invariant-mass distributions in the sextet-favored reactions (a) $\Upsilon\to\bar\Xi_c'\Lambda(1520)D\eta$ and (b) $\Xi_b'\to\pi^-\Lambda(1520)D\eta$. The vertical dashed line indicates the $D_s\bar K$ threshold.}
\label{fig:deta}
\end{minipage}\hfill
\begin{minipage}[t]{0.48\textwidth}
\centering
\includegraphics[width=0.88\linewidth]{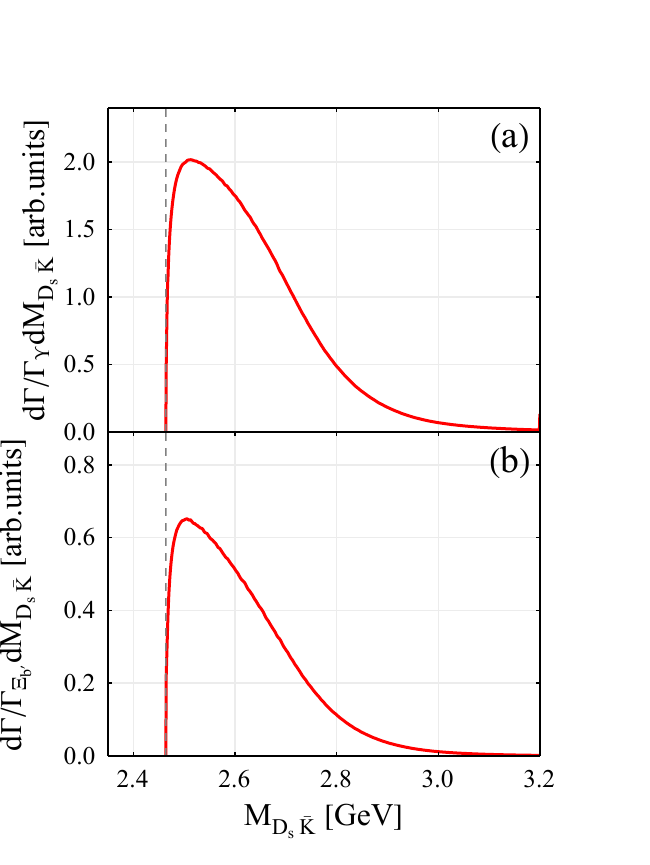}
\caption{Normalized $D_s\bar K$ invariant-mass distributions in the sextet-favored reactions (a) $\Upsilon\to\bar\Xi_c'\Lambda(1520)D_s\bar K$ and (b) $\Xi_b'\to\pi^-\Lambda(1520)D_s\bar K$.}
\label{fig:dsk}
\end{minipage}
\end{figure*}

The coupled-channel interpretation makes correlated predictions for the other two final states, as shown in Figs.~\ref{fig:deta} and \ref{fig:dsk}. In the $D\eta$ spectra, the strength that develops above the $D\eta$ threshold culminates in a sharp cusp when the $D_s\bar K$ channel opens. In the $D_s\bar K$ spectra, the proximity of the higher pole to the threshold produces an asymmetric near-threshold enhancement rather than an isolated resonance peak. The strong and weak reactions again yield the same qualitative behavior. Observing these correlated structures at common mass scales in the $D\pi$, $D\eta$, and $D_s\bar K$ channels would provide substantially stronger evidence for the coupled-channel pole than a bump in the $D\pi$ spectrum alone and would enable a simultaneous pole analysis of all three channels.

\begin{figure*}[t]
\begin{minipage}[t]{0.48\textwidth}
\centering
\includegraphics[width=0.88\linewidth]{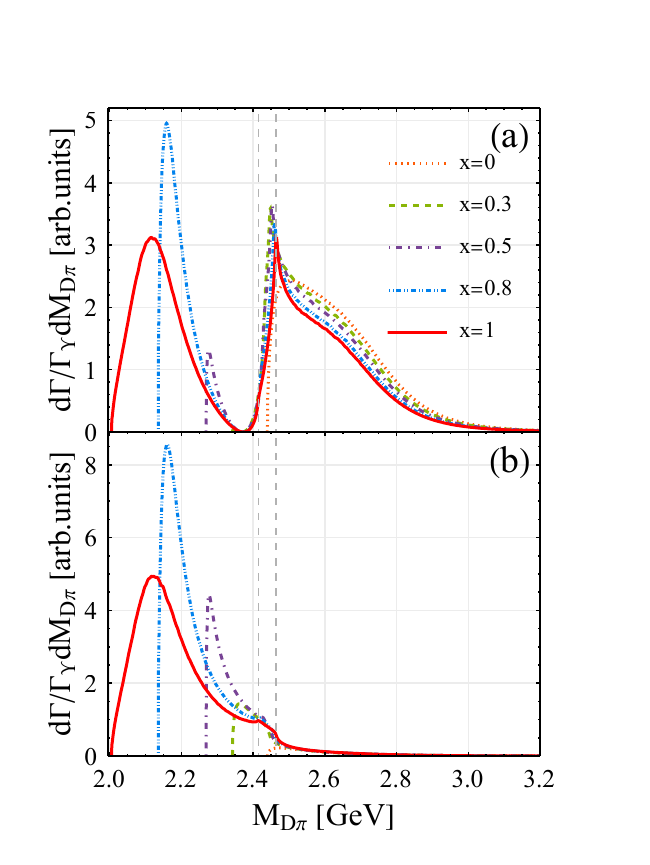}
\caption{$D\pi$ invariant-mass distributions in (a) $\Upsilon\to\bar\Xi_c'\Lambda(1520)D\pi$ and (b) $\Upsilon\to\bar\Xi_c\Lambda(1520)D\pi$ for different values of the $SU(3)$-breaking parameter $x$ in the ideal flavor basis. The dashed vertical lines mark the physical $D\eta$ and $D_s\bar K$ thresholds.}
\label{fig:strong-su3}
\end{minipage}\hfill
\begin{minipage}[t]{0.48\textwidth}
\centering
\includegraphics[width=0.88\linewidth]{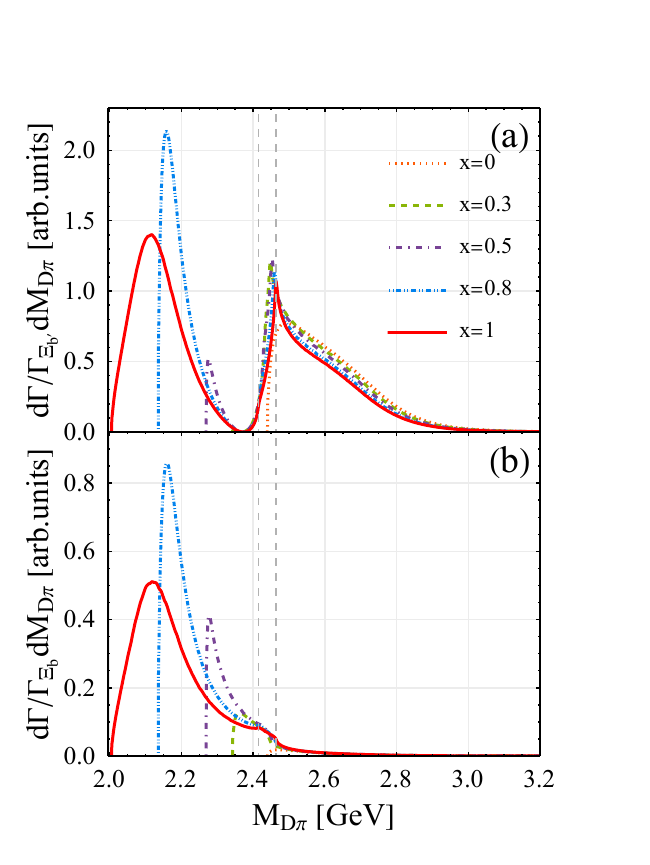}
\caption{$D\pi$ invariant-mass distributions in (a) $\Xi_b'\to\pi^-\Lambda(1520)D\pi$ and (b) $\Xi_b\to\pi^-\Lambda(1520)D\pi$ for different values of the $SU(3)$-breaking parameter $x$ in the ideal flavor basis. The dashed vertical lines mark the physical $D\eta$ and $D_s\bar K$ thresholds.}
\label{fig:weak-su3}
\end{minipage}
\end{figure*}

The evolution shown in Figs.~\ref{fig:strong-su3} and \ref{fig:weak-su3} verifies this interpretation and explains why the lower pole is visible in a sextet-tagged reaction. At the symmetric point ($x=0$), the three thresholds coincide, the coupled-channel amplitude remains diagonal in the irreducible-representation basis, and the source selection is exact. The sextet source in panel (a) of each figure therefore produces the strong threshold enhancement associated with the sextet virtual state without a contribution from the antitriplet bound state. Conversely, although the antitriplet source in panel (b) couples to the lower bound state, that pole lies below the common threshold and cannot appear as an above-threshold resonance peak.

As $x$ increases, the thresholds separate and the poles evolve through the relevant Riemann sheets. The channel-dependent loop functions and interaction kernels no longer preserve a strictly diagonal sextet--antitriplet decomposition. At the same time, the $D\pi$ threshold moves below the lower pole, allowing the latter to generate the broad low-mass enhancement and, through representation mixing, to become visible in both source channels. The higher pole evolves into the enhancement between the $D\eta$ and $D_s\bar K$ thresholds and the cusp at the $D_s\bar K$ opening. By $x=1$, these continuously connected features reproduce the physical spectra in Figs.~\ref{fig:strongdpi} and \ref{fig:weakdpi}. The evolution demonstrates that the two mass regions arise from poles connected to different $SU(3)$ representations, while also making clear that the physical flavor filter is an approximate pole selector rather than an exact projector.

The most discriminating flavor-filter comparison would combine the paired tags, $\bar\Xi_c'$ versus $\bar\Xi_c$ or $\Xi_b'$ versus $\Xi_b$, with a simultaneous analysis of the $D\pi$, $D\eta$, and $D_s\bar K$ spectra. Possible octet admixtures in the $\Lambda(1520)$, $\Xi_Q$--$\Xi_Q'$ mixing, explicit $SU(3)$ breaking in the short-distance vertices, and nonresonant production mechanisms can dilute the idealized contrast and should be included in a precision fit. Such effects mainly introduce additional source components and change the normalizations; if their energy dependence is smooth, they are unlikely to mimic both the representation-dependent redistribution of strength and the correlated threshold structures in all three channels. A quantitative extraction should consequently employ a coupled-channel amplitude with complex production coefficients, while the qualitative line-shape pattern identified here remains the robust prediction.

An important experimental qualification concerns the weak sextet-tagged channel. The physical $\Xi_b'$ de-excites predominantly through strong or electromagnetic interactions, so a decay driven by the much slower $b$-quark weak transition is expected to have a very small branching fraction~\cite{ParticleDataGroup:2024cfk}. The $\Xi_b'$ spectra should therefore be viewed primarily as an idealized demonstration of the sextet filter, whereas the $\Upsilon\to\bar\Xi_c'\Lambda(1520)\mathcal{D}\Phi$ modes provide the more realistic sextet-tagged target. In the latter case, the final-state $\bar\Xi_c'$ can be reconstructed through its radiative decay.

\section{Summary}\label{sec:summary}

We have proposed a flavor-filter strategy for resolving the two-pole structure associated with the $D_0^*(2300)$. The method exploits the predominantly flavor-singlet character of the $\Lambda(1520)$ and the different $SU(3)$ representations of $\Xi_Q$ and $\Xi_Q'$. In the strong decays $\Upsilon\to\bar\Xi_c^{(\prime)}\Lambda(1520)\mathcal{D}\Phi$ and the weak decays $\Xi_b^{(\prime)}\to\pi^-\Lambda(1520)\mathcal{D}\Phi$, these flavor assignments generate approximately antitriplet- and sextet-selective $\mathcal{D}\Phi$ sources.

The calculated spectra exhibit complementary signatures. Sextet-favored reactions prominently expose the higher pole near $2.45~\mathrm{GeV}$ through an enhancement between the $D\eta$ and $D_s\bar K$ thresholds and a cusp at the $D_s\bar K$ opening, while $SU(3)$-breaking rescattering also allows the lower antitriplet pole to remain visible. Conversely, antitriplet-favored reactions are dominated by the lower pole near $2.1~\mathrm{GeV}$ and show only a much weaker higher-energy structure. This representation-dependent redistribution of spectral strength is reproduced in both strong and weak production and evolves continuously from the antitriplet bound state and sextet virtual state at the symmetric point.

Absolute branching fractions cannot yet be predicted because the short-distance production couplings are unknown. Moreover, the rapid strong or electromagnetic de-excitation of the physical $\Xi_b'$ makes its weak decay an idealized rather than a near-term experimental realization of the sextet filter. The $\Upsilon$ modes thus provide the cleaner experimental target, while the weak-decay calculation illustrates the robustness of the predicted line-shape pattern against smooth changes in the production mechanism. Nevertheless, the proposed test relies primarily on correlated line-shape features across the $D\pi$, $D\eta$, and $D_s\bar K$ channels. Measurements of the accessible modes would directly test the two-pole interpretation and could provide clear invariant-mass evidence for a charmed-meson resonance with a dominant flavor-sextet component.

% Insert the appropriate funding information and acknowledgments before submission.

\appendix
\section{Parameters of the unitarized amplitude}\label{sec:appendix}

For completeness, we summarize the interaction used to generate the $\mathcal{D}\Phi$ scattering amplitude. The unitarized matrix is given by Eq.~\eqref{eq:unitarized}, with $V=V_{\rm LO}+V_{\rm NLO}$. In dimensional regularization, the scalar loop function for channel $i$ is
\begin{align}
G_i(s)=\frac{1}{16\pi^2}\bigg\{&\tilde a(\lambda)+\ln\frac{m_2^2}{\lambda^2}
+\frac{m_1^2-m_2^2+s}{2s}\ln\frac{m_1^2}{m_2^2}\notag\\
&+\frac{\sigma}{2s}\,\mathcal L(s)\bigg\},
\label{eq:loopfunction}\\
\mathcal L(s)={}&\ln(s-m_1^2+m_2^2+\sigma)\notag\\
&-\ln(-s+m_1^2-m_2^2+\sigma)\notag\\
&+\ln(s+m_1^2-m_2^2+\sigma)\notag\\
&-\ln(-s-m_1^2+m_2^2+\sigma),
\label{eq:loop-logarithms}\\
\sigma={}&\sqrt{\lambda_K(s,m_1^2,m_2^2)},
\label{eq:loop-sigma}\\
\lambda_K(a,b,c)={}&a^2+b^2+c^2-2ab-2ac-2bc.
\label{eq:kallen-function}
\end{align}
Here $m_1$ and $m_2$ are the two meson masses in channel $i$, $\lambda$ is the dimensional-regularization scale, $\tilde a(\lambda)$ is the subtraction constant, and the auxiliary quantities $\mathcal L(s)$, $\sigma$, and the K\"all\'en function $\lambda_K$ are defined in Eqs.~\eqref{eq:loop-logarithms}--\eqref{eq:kallen-function}. We take $\lambda=1~\mathrm{GeV}$ and $\tilde a(\lambda)=-1.88$, as in Ref.~\cite{Liu:2012zya}. Alternative determinations of the subtraction prescription and NLO low-energy constants from scattering lengths and lattice spectra can be found in Refs.~\cite{Guo:2009ct,Altenbuchinger:2013vwa,Yao:2015qia,Du:2016puz,Lutz:2022dcl}. Before $S$-wave projection, the interaction kernel is
\begin{align}
V(s,t,u)=\frac{1}{F^2}\bigg[&\frac{C_{\rm LO}}{4}(s-u)-4C_0h_0+2C_1h_1\notag\\
&-2C_{24}H_{24}(s,t,u)+2C_{35}H_{35}(s,t,u)\bigg],
\label{eq:potential-appendix}
\end{align}
The momentum-dependent structures entering this kernel are defined by
\begin{align}
&H_{24}(s,t,u)=2h_{24}\,p_2\!\cdot p_4+h_4\mathcal P,
\label{eq:H24}\\
&H_{35}(s,t,u)=h_{35}\,p_2\!\cdot p_4+h_5\mathcal P.
\label{eq:H35}\\
&\mathcal P = (p_1\!\cdot p_2)(p_3\!\cdot p_4)
+(p_1\!\cdot p_4)(p_2\!\cdot p_3)
-2\bar M_D^2(p_2\!\cdot p_4),
\label{eq:momentum-polynomial}
\end{align}
For the process $\mathcal{D}(p_1)\Phi(p_2)\to\mathcal{D}(p_3)\Phi(p_4)$, $p_1$ and $p_2$ ($p_3$ and $p_4$) are the incoming (outgoing) four-momenta. We use $F=92.4~\mathrm{MeV}$ and $\bar M_D=(M_D+M_{D_s})/2$. The channel coefficients and low-energy constants are listed in Tables~\ref{tab:Ccoefficients} and \ref{tab:hcoefficients}; only independent transitions are shown, and the reverse coefficients follow from time-reversal invariance.

\begin{table*}[t]
\centering
\caption{Channel coefficients in the $(S,I)=(0,1/2)$ sector.}
\label{tab:Ccoefficients}
\begin{tabular}{lccccc}
\toprule
Channel & $C_{\rm LO}$ & $C_0$ & $C_1$ & $C_{24}$ & $C_{35}$\\
\midrule
$D\pi\to D\pi$ & $-2$ & $M_\pi^2$ & $-M_\pi^2$ & $1$ & $1$\\
$D\eta\to D\eta$ & $0$ & $M_\eta^2$ & $-M_\pi^2/3$ & $1$ & $1/3$\\
$D_s\bar K\to D_s\bar K$ & $-1$ & $M_K^2$ & $-M_K^2$ & $1$ & $1$\\
$D\eta\to D\pi$ & $0$ & $0$ & $-M_\pi^2$ & $0$ & $1$\\
$D_s\bar K\to D\pi$ & $-\sqrt6/2$ & $0$ & $-\sqrt6(M_K^2+M_\pi^2)/4$ & $0$ & $\sqrt6/2$\\
$D_s\bar K\to D\eta$ & $-\sqrt6/2$ & $0$ & $\sqrt6(5M_K^2-3M_\pi^2)/12$ & $0$ & $-1/\sqrt6$\\
\bottomrule
\end{tabular}
\end{table*}

\begin{table}[tb]
\centering
\caption{Low-energy constants used in the NLO interaction.}
\label{tab:hcoefficients}
\begin{tabular}{cccccc}
\toprule
$h_0$ & $h_1$ & $h_{24}$ & $h_4$ & $h_{35}$ & $h_5$\\
\midrule
$0.014$ & $0.42$ & $-0.10$ & $-0.30/\bar M_D^2$ & $0.26$ & $-1.94/\bar M_D^2$\\
\bottomrule
\end{tabular}
\end{table}

\bibliography{refs}

@article{ParticleDataGroup:2024cfk,
    author = "Navas, S. and others",
    collaboration = "Particle Data Group",
    title = "{Review of particle physics}",
    doi = "10.1103/PhysRevD.110.030001",
    journal = "Phys. Rev. D",
    volume = "110",
    number = "3",
    pages = "030001",
    year = "2024"
}

@article{Guo:2025vjw,
    author = "Guo, Feng-Kun and Zou, Bing-Song",
    title = "{Evidence for the existence of a flavor-sextet charmed meson?}",
    eprint = "2504.21442",
    archivePrefix = "arXiv",
    primaryClass = "hep-ph",
    doi = "10.1016/j.scib.2025.06.004",
    journal = "Sci. Bull.",
    volume = "70",
    pages = "2535--2537",
    year = "2025"
}

@article{BaBar:2003oey,
    author = "Aubert, B. and others",
    collaboration = "BABAR",
    title = "{Observation of a narrow meson decaying to $D_s^+\pi^0$ at a mass of $2.32~\mathrm{GeV}/c^2$}",
    eprint = "hep-ex/0304021",
    archivePrefix = "arXiv",
    reportNumber = "SLAC-PUB-9711",
    doi = "10.1103/PhysRevLett.90.242001",
    journal = "Phys. Rev. Lett.",
    volume = "90",
    pages = "242001",
    year = "2003"
}

@article{CLEO:2003ggt,
    author = "Besson, D. and others",
    collaboration = "CLEO",
    title = "{Observation of a narrow resonance of mass $2.46~\mathrm{GeV}/c^2$ decaying to $D_s^{*+}\pi^0$ and confirmation of the $D_{sJ}^*(2317)$ state}",
    eprint = "hep-ex/0305100",
    archivePrefix = "arXiv",
    reportNumber = "CLNS-03-1826, CLEO-03-09",
    doi = "10.1103/PhysRevD.68.032002",
    journal = "Phys. Rev. D",
    volume = "68",
    pages = "032002",
    year = "2003",
    note = "[Erratum: Phys. Rev. D 75, 119908 (2007)]"
}

@article{Belle:2003nnu,
    author = "Abe, K. and others",
    collaboration = "Belle",
    title = "{Study of $B^-\to D^{**0}\pi^-$ ($D^{**0}\to D^{(*)+}\pi^-$) decays}",
    eprint = "hep-ex/0307021",
    archivePrefix = "arXiv",
    doi = "10.1103/PhysRevD.69.112002",
    journal = "Phys. Rev. D",
    volume = "69",
    pages = "112002",
    year = "2004"
}

@article{Belle:2003drx,
    author = "Krokovny, P. and others",
    collaboration = "Belle",
    title = "{Observation of the $D_{sJ}(2317)$ and $D_{sJ}(2457)$ in $B$ decays}",
    eprint = "hep-ex/0308019",
    archivePrefix = "arXiv",
    doi = "10.1103/PhysRevLett.91.262002",
    journal = "Phys. Rev. Lett.",
    volume = "91",
    pages = "262002",
    year = "2003"
}

@article{LHCb:2016lxy,
    author = "Aaij, R. and others",
    collaboration = "LHCb",
    title = "{Amplitude analysis of $B^-\to D^+\pi^-\pi^-$ decays}",
    eprint = "1608.01289",
    archivePrefix = "arXiv",
    primaryClass = "hep-ex",
    reportNumber = "CERN-EP-2016-184, LHCb-PAPER-2016-026",
    doi = "10.1103/PhysRevD.94.072001",
    journal = "Phys. Rev. D",
    volume = "94",
    number = "7",
    pages = "072001",
    year = "2016"
}

@article{Bardeen:2003kt,
    author = "Bardeen, William A. and Eichten, Estia J. and Hill, Christopher T.",
    title = "{Chiral multiplets of heavy-light mesons}",
    eprint = "hep-ph/0305049",
    archivePrefix = "arXiv",
    reportNumber = "FERMILAB-PUB-03-071-T",
    doi = "10.1103/PhysRevD.68.054024",
    journal = "Phys. Rev. D",
    volume = "68",
    pages = "054024",
    year = "2003"
}

@article{Barnes:2003dj,
    author = "Barnes, T. and Close, F. E. and Lipkin, H. J.",
    title = "{Implications of a $DK$ molecule at $2.32~\mathrm{GeV}$}",
    eprint = "hep-ph/0305025",
    archivePrefix = "arXiv",
    doi = "10.1103/PhysRevD.68.054006",
    journal = "Phys. Rev. D",
    volume = "68",
    pages = "054006",
    year = "2003"
}

@article{Cleven:2010aw,
    author = "Cleven, Martin and Guo, Feng-Kun and Hanhart, Christoph and Mei{\ss}ner, Ulf-G.",
    title = "{Light meson mass dependence of the positive-parity heavy-strange mesons}",
    eprint = "1009.3804",
    archivePrefix = "arXiv",
    primaryClass = "hep-ph",
    doi = "10.1140/epja/i2011-11019-2",
    journal = "Eur. Phys. J. A",
    volume = "47",
    pages = "19",
    year = "2011"
}

@article{Mohler:2013rwa,
    author = "Mohler, Daniel and Lang, C. B. and Leskovec, Luka and Prelovsek, Sasa and Woloshyn, R. M.",
    title = "{$D_{s0}^*(2317)$ meson and $D$-meson--kaon scattering from lattice QCD}",
    eprint = "1308.3175",
    archivePrefix = "arXiv",
    primaryClass = "hep-lat",
    reportNumber = "FERMILAB-PUB-13-318-T",
    doi = "10.1103/PhysRevLett.111.222001",
    journal = "Phys. Rev. Lett.",
    volume = "111",
    number = "22",
    pages = "222001",
    year = "2013"
}

@article{Lang:2014yfa,
    author = "Lang, C. B. and Leskovec, Luka and Mohler, Daniel and Prelovsek, Sasa and Woloshyn, R. M.",
    title = "{$D_s$ mesons with $DK$ and $D^*K$ scattering near threshold}",
    eprint = "1403.8103",
    archivePrefix = "arXiv",
    primaryClass = "hep-lat",
    reportNumber = "FERMILAB-PUB-14-063-T",
    doi = "10.1103/PhysRevD.90.034510",
    journal = "Phys. Rev. D",
    volume = "90",
    number = "3",
    pages = "034510",
    year = "2014"
}

@article{MartinezTorres:2014son,
    author = "Mart{\'\i}nez Torres, A. and Oset, E. and Prelovsek, S. and Ramos, A.",
    title = "{Reanalysis of lattice-QCD spectra leading to the $D_{s0}^*(2317)$ and $D_{s1}(2460)$}",
    eprint = "1412.1706",
    archivePrefix = "arXiv",
    primaryClass = "hep-lat",
    doi = "10.1007/JHEP05(2015)153",
    journal = "JHEP",
    volume = "05",
    pages = "153",
    year = "2015"
}

@article{Bali:2017pdv,
    author = "Bali, Gunnar S. and Collins, Sara and Cox, Antonio and Sch{\"a}fer, Andreas",
    title = "{Masses and decay constants of the $D_{s0}^*(2317)$ and $D_{s1}(2460)$ from $N_f=2$ lattice QCD close to the physical point}",
    eprint = "1706.01247",
    archivePrefix = "arXiv",
    primaryClass = "hep-lat",
    doi = "10.1103/PhysRevD.96.074501",
    journal = "Phys. Rev. D",
    volume = "96",
    number = "7",
    pages = "074501",
    year = "2017"
}

@article{Albaladejo:2016lbb,
    author = "Albaladejo, Miguel and Fernandez-Soler, Pedro and Guo, Feng-Kun and Nieves, Juan",
    title = "{Two-pole structure of the $D^\ast_0(2400)$}",
    eprint = "1610.06727",
    archivePrefix = "arXiv",
    primaryClass = "hep-ph",
    doi = "10.1016/j.physletb.2017.02.036",
    journal = "Phys. Lett. B",
    volume = "767",
    pages = "465--469",
    year = "2017"
}

@article{Meng:2022ozq,
    author = "Meng, Lu and Wang, Bo and Wang, Guang-Juan and Zhu, Shi-Lin",
    title = "{Chiral perturbation theory for heavy hadrons and chiral effective field theory for heavy hadronic molecules}",
    eprint = "2204.08716",
    archivePrefix = "arXiv",
    primaryClass = "hep-ph",
    doi = "10.1016/j.physrep.2023.04.003",
    journal = "Phys. Rept.",
    volume = "1019",
    pages = "1--149",
    year = "2023"
}

@article{Moir:2016srx,
    author = "Moir, Graham and Peardon, Michael and Ryan, Sin{\'e}ad M. and Thomas, Christopher E. and Wilson, David J.",
    title = "{Coupled-Channel $D\pi$, $D\eta$ and $D_{s}\bar{K}$ Scattering from Lattice QCD}",
    eprint = "1607.07093",
    archivePrefix = "arXiv",
    primaryClass = "hep-lat",
    reportNumber = "DAMTP-2016-48",
    doi = "10.1007/JHEP10(2016)011",
    journal = "JHEP",
    volume = "10",
    pages = "011",
    year = "2016"
}

@article{Yeo:2024chk,
    author = "Yeo, J. Daniel E. and Thomas, Christopher E. and Wilson, David J.",
    collaboration = "Hadron Spectrum",
    title = "{DK/D{\ensuremath{\pi}} scattering and an exotic virtual bound state at the SU(3) flavour symmetric point from lattice QCD}",
    eprint = "2403.10498",
    archivePrefix = "arXiv",
    primaryClass = "hep-lat",
    doi = "10.1007/JHEP07(2024)012",
    journal = "JHEP",
    volume = "07",
    pages = "012",
    year = "2024"
}

@article{Du:2017zvv,
    author = "Du, Meng-Lin and Albaladejo, Miguel and Fern{\'a}ndez-Soler, Pedro and Guo, Feng-Kun and Hanhart, Christoph and Mei{\ss}ner, Ulf-G. and Nieves, Juan and Yao, De-Liang",
    title = "{Towards a new paradigm for heavy-light meson spectroscopy}",
    eprint = "1712.07957",
    archivePrefix = "arXiv",
    primaryClass = "hep-ph",
    doi = "10.1103/PhysRevD.98.094018",
    journal = "Phys. Rev. D",
    volume = "98",
    number = "9",
    pages = "094018",
    year = "2018"
}

@article{Asokan:2022usm,
    author = "Asokan, Anuvind and Tang, Meng-Na and Guo, Feng-Kun and Hanhart, Christoph and Kamiya, Yuki and Mei{\ss}ner, Ulf-G.",
    title = "{Can the two-pole structure of the $D_{0}^{*}(2300)$ be understood from recent lattice data?}",
    eprint = "2212.07856",
    archivePrefix = "arXiv",
    primaryClass = "hep-ph",
    doi = "10.1140/epjc/s10052-023-11953-6",
    journal = "Eur. Phys. J. C",
    volume = "83",
    number = "9",
    pages = "850",
    year = "2023"
}

@article{LHCb:2024iuo,
    author = "Aaij, R. and others",
    collaboration = "LHCb",
    title = "{Study of $D_{s1}(2460)^{+}\to D_{s}^{+}\pi^{+}\pi^{-}$ in $B\to {\bar{D}}^{(*)}D_{s}^{+}\pi^{+}\pi^{-}$ decays}",
    eprint = "2411.03399",
    archivePrefix = "arXiv",
    primaryClass = "hep-ex",
    reportNumber = "LHCb-PAPER-2024-033, CERN-EP-2024-264",
    doi = "10.1016/j.scib.2025.02.025",
    journal = "Sci. Bull.",
    volume = "70",
    pages = "1432--1444",
    year = "2025"
}

@article{Maiani:2024quj,
    author = "Maiani, L. and Polosa, A. D. and Riquer, V.",
    title = "{Open charm tetraquarks in broken SU(3)F symmetry}",
    eprint = "2405.08545",
    archivePrefix = "arXiv",
    primaryClass = "hep-ph",
    doi = "10.1103/PhysRevD.110.034014",
    journal = "Phys. Rev. D",
    volume = "110",
    number = "3",
    pages = "034014",
    year = "2024"
}

@article{Du:2025beb,
    author = "Du, M. -L. and Guo, F. -K. and Hanhart, C. and Herren, F. and Kubis, B. and van Tonder, R.",
    title = "{Discovering the $D_0^\ast(2100)$ in $B$ semileptonic decays}",
    eprint = "2509.12133",
    archivePrefix = "arXiv",
    primaryClass = "hep-ph",
    reportNumber = "ZU-TH 57/25",
    month = "9",
    year = "2025",
    journal = "arXiv preprint"
}

@article{He:2024uau,
    author = "He, Ying-Bo and Liu, Xiao-Hai and Geng, Li-Sheng and Guo, Feng-Kun and Xie, Ju-Jun",
    title = "{Identifying the two-pole structure of the $\Lambda(1405)$ using an SU(3) flavor filter}",
    eprint = "2407.13486",
    archivePrefix = "arXiv",
    primaryClass = "hep-ph",
    doi = "10.1103/5sr9-tzj6",
    journal = "Phys. Rev. D",
    volume = "113",
    number = "5",
    pages = "L051501",
    year = "2026"
}

@article{Oller:2000fj,
    author = "Oller, J. A. and Meissner, Ulf-G.",
    title = "{Chiral dynamics in the presence of bound states: Kaon nucleon interactions revisited}",
    eprint = "hep-ph/0011146",
    archivePrefix = "arXiv",
    doi = "10.1016/S0370-2693(01)00078-8",
    journal = "Phys. Lett. B",
    volume = "500",
    pages = "263--272",
    year = "2001"
}

@article{Jido:2003cb,
    author = "Jido, Daisuke and Oller, J. A. and Oset, Eulogio and Ramos, A. and Meissner, Ulf-G.",
    title = "{Chiral dynamics of the two $\Lambda(1405)$ states}",
    eprint = "nucl-th/0303062",
    archivePrefix = "arXiv",
    doi = "10.1016/S0375-9474(03)01598-7",
    journal = "Nucl. Phys. A",
    volume = "725",
    pages = "181--200",
    year = "2003"
}

@article{Guo:2023mku,
    author = "Guo, Feng-Kun and Kamiya, Yuki and Mai, Maxim and Meissner, Ulf-G.",
    title = "{New insights into the nature of the $\Lambda(1380)$ and $\Lambda(1405)$ resonances away from the SU(3) limit}",
    eprint = "2308.07658",
    archivePrefix = "arXiv",
    primaryClass = "hep-ph",
    doi = "10.1016/j.physletb.2023.138264",
    journal = "Phys. Lett. B",
    volume = "846",
    pages = "138264",
    year = "2023"
}

@article{Bulava:2023uma,
    author = "Bulava, John and Cid-Mora, Barbara and Hanlon, Andrew D. and Horz, Ben and Mohler, Daniel and Morningstar, Colin and Moscoso, Joseph and Nicholson, Amy and Romero-Lopez, Fernando and Skinner, Sarah and Walker-Loud, Andre",
    collaboration = "Baryon Scattering",
    title = "{Two-pole nature of the $\Lambda(1405)$ resonance from lattice QCD}",
    eprint = "2307.13471",
    archivePrefix = "arXiv",
    primaryClass = "hep-lat",
    doi = "10.1103/PhysRevD.109.014511",
    journal = "Phys. Rev. D",
    volume = "109",
    number = "1",
    pages = "014511",
    year = "2024"
}

@article{SuarezSucunza:2026lambda,
    author = "Suarez Sucunza, Javier and Luu, Thomas and Urbach, Carsten",
    title = "{The $\Lambda(1405)$ at the SU(3) point in lattice QCD}",
    eprint = "2602.19928",
    archivePrefix = "arXiv",
    primaryClass = "hep-lat",
    month = "2",
    year = "2026",
    journal = "arXiv preprint"
}

@article{Xie:2023two,
    author = "Xie, Jia-Ming and Lu, Jun-Xu and Geng, Li-Sheng and Zou, Bing-Song",
    title = "{Two-pole structures as a universal phenomenon dictated by coupled-channel chiral dynamics}",
    eprint = "2307.11631",
    archivePrefix = "arXiv",
    primaryClass = "hep-ph",
    doi = "10.1103/PhysRevD.108.L111502",
    journal = "Phys. Rev. D",
    volume = "108",
    number = "11",
    pages = "L111502",
    year = "2023"
}

@article{Liu:2012zya,
    author = "Liu, Liuming and Orginos, Kostas and Guo, Feng-Kun and Hanhart, Christoph and Meissner, Ulf-G.",
    title = "{Interactions of charmed mesons with light pseudoscalar mesons from lattice QCD and implications on the nature of the $D_{s0}^*(2317)$}",
    eprint = "1208.4535",
    archivePrefix = "arXiv",
    primaryClass = "hep-lat",
    reportNumber = "JLAB-THY-12-1599",
    doi = "10.1103/PhysRevD.87.014508",
    journal = "Phys. Rev. D",
    volume = "87",
    number = "1",
    pages = "014508",
    year = "2013"
}

@article{Guo:2008gp,
    author = "Guo, Feng-Kun and Hanhart, Christoph and Krewald, Siegfried and Meissner, Ulf-G.",
    title = "{Subleading contributions to the width of the D*(s0)(2317)}",
    eprint = "0806.3374",
    archivePrefix = "arXiv",
    primaryClass = "hep-ph",
    reportNumber = "FZJ-IKP(TH)-2008-09, HISKP-TH-08-09",
    doi = "10.1016/j.physletb.2008.07.060",
    journal = "Phys. Lett. B",
    volume = "666",
    pages = "251--255",
    year = "2008"
}

@article{Klempt:2020bdu,
    author = "Klempt, E. and Burkert, V. and Thoma, U. and Tiator, L. and Workman, R.",
    collaboration = "Baryon@PDG Group",
    title = "{$\varLambda $ and $\varSigma $ Excitations and the Quark Model}",
    eprint = "2007.04232",
    archivePrefix = "arXiv",
    primaryClass = "nucl-ex",
    doi = "10.1140/epja/s10050-020-00261-2",
    journal = "Eur. Phys. J. A",
    volume = "56",
    number = "10",
    pages = "261",
    year = "2020"
}

@article{Kamano:2015hxa,
    author = "Kamano, H. and Nakamura, S. X. and Lee, T. -S. H. and Sato, T.",
    title = "{Dynamical coupled-channels model of $K^- p$ reactions. II. Extraction of $\Lambda^*$ and $\Sigma^*$ hyperon resonances}",
    eprint = "1506.01768",
    archivePrefix = "arXiv",
    primaryClass = "nucl-th",
    doi = "10.1103/PhysRevC.92.025205",
    journal = "Phys. Rev. C",
    volume = "92",
    number = "2",
    pages = "025205",
    year = "2015",
    note = "[Erratum: Phys.Rev.C 95, 049903 (2017)]"
}

@article{Zhang:2013sva,
    author = "Zhang, H. and Tulpan, J. and Shrestha, M. and Manley, D. M.",
    title = "{Multichannel parametrization of $\bar K N$ scattering amplitudes and extraction of resonance parameters}",
    eprint = "1305.4575",
    archivePrefix = "arXiv",
    primaryClass = "hep-ph",
    doi = "10.1103/PhysRevC.88.035205",
    journal = "Phys. Rev. C",
    volume = "88",
    number = "3",
    pages = "035205",
    year = "2013"
}

@article{Matveev:2019igl,
    author = "Matveev, M. and Sarantsev, A. V. and Nikonov, V. A. and Anisovich, A. V. and Thoma, U. and Klempt, E.",
    title = "{Hyperon I: Partial-wave amplitudes for K$^{-}$p scattering}",
    eprint = "1907.03645",
    archivePrefix = "arXiv",
    primaryClass = "nucl-ex",
    doi = "10.1140/epja/i2019-12878-y",
    journal = "Eur. Phys. J. A",
    volume = "55",
    number = "10",
    pages = "179",
    year = "2019"
}

@article{Sarantsev:2019xxm,
    author = "Sarantsev, A. V. and Matveev, M. and Nikonov, V. A. and Anisovich, A. V. and Thoma, U. and Klempt, E.",
    title = "{Hyperon II: Properties of excited hyperons}",
    eprint = "1907.13387",
    archivePrefix = "arXiv",
    primaryClass = "nucl-ex",
    doi = "10.1140/epja/i2019-12880-5",
    journal = "Eur. Phys. J. A",
    volume = "55",
    number = "10",
    pages = "180",
    year = "2019"
}

@article{Cheng:2006dk,
    author = "Cheng, Hai-Yang and Chua, Chun-Khiang",
    title = "{Strong Decays of Charmed Baryons in Heavy Hadron Chiral Perturbation Theory}",
    eprint = "hep-ph/0610283",
    archivePrefix = "arXiv",
    doi = "10.1103/PhysRevD.75.014006",
    journal = "Phys. Rev. D",
    volume = "75",
    pages = "014006",
    year = "2007"
}

@article{Ke:2024aux,
    author = "Ke, Hong-Wei and Fang, Gang-Yang and Shi, Yan-Liang",
    title = "{Study on the mixing of {\ensuremath{\Xi}}c and {\ensuremath{\Xi}}c' by the transition {\ensuremath{\Xi}}b{\textrightarrow}{\ensuremath{\Xi}}c(')}",
    eprint = "2401.11106",
    archivePrefix = "arXiv",
    primaryClass = "hep-ph",
    doi = "10.1103/PhysRevD.109.073006",
    journal = "Phys. Rev. D",
    volume = "109",
    number = "7",
    pages = "073006",
    year = "2024"
}

@article{Liu:2023feb,
    author = "Liu, Hang and Liu, Liuming and Sun, Peng and Sun, Wei and Tan, Jin-Xin and Wang, Wei and Yang, Yi-Bo and Zhang, Qi-An",
    title = "{{\ensuremath{\Xi}}c{\ensuremath{-}}{\ensuremath{\Xi}}c' mixing from lattice QCD}",
    eprint = "2303.17865",
    archivePrefix = "arXiv",
    primaryClass = "hep-lat",
    doi = "10.1016/j.physletb.2023.137941",
    journal = "Phys. Lett. B",
    volume = "841",
    pages = "137941",
    year = "2023"
}

@article{Sun:2023noo,
    author = "Sun, Xiao-Yu and Zhang, Fu-Wei and Shi, Yu-Ji and Zhao, Zhen-Xing",
    title = "{Revisiting $\Xi _{Q}-\Xi _{Q}^{\prime }$ mixing in QCD sum rules}",
    eprint = "2305.08050",
    archivePrefix = "arXiv",
    primaryClass = "hep-ph",
    doi = "10.1140/epjc/s10052-023-12042-4",
    journal = "Eur. Phys. J. C",
    volume = "83",
    number = "10",
    pages = "961",
    year = "2023"
}

@article{Aliev:2010ra,
    author = "Aliev, T. M. and Ozpineci, A. and Zamiralov, V.",
    title = "{Mixing Angle of Hadrons in QCD: A New View}",
    eprint = "1007.0814",
    archivePrefix = "arXiv",
    primaryClass = "hep-ph",
    doi = "10.1103/PhysRevD.83.016008",
    journal = "Phys. Rev. D",
    volume = "83",
    pages = "016008",
    year = "2011"
}

@article{Schat:2001xr,
    author = "Schat, C. L. and Goity, J. L. and Scoccola, N. N.",
    title = "{Masses of the 70- baryons in large N(c) QCD}",
    eprint = "hep-ph/0111082",
    archivePrefix = "arXiv",
    reportNumber = "JLAB-THY-01-35",
    doi = "10.1103/PhysRevLett.88.102002",
    journal = "Phys. Rev. Lett.",
    volume = "88",
    pages = "102002",
    year = "2002"
}

@article{Melde:2008yr,
    author = "Melde, T. and Plessas, Willibald and Sengl, B.",
    title = "{Quark-Model Identification of Baryon Ground and Resonant States}",
    eprint = "0806.1454",
    archivePrefix = "arXiv",
    primaryClass = "hep-ph",
    doi = "10.1103/PhysRevD.77.114002",
    journal = "Phys. Rev. D",
    volume = "77",
    pages = "114002",
    year = "2008"
}

@article{Roca:2005nm,
    author = "Roca, L. and Oset, E. and Singh, J.",
    title = "{Low lying axial-vector mesons as dynamically generated resonances}",
    eprint = "hep-ph/0503273",
    archivePrefix = "arXiv",
    doi = "10.1103/PhysRevD.72.014002",
    journal = "Phys. Rev. D",
    volume = "72",
    pages = "014002",
    year = "2005"
}

@article{Geng:2006yb,
    author = "Geng, L. S. and Oset, E. and Roca, L. and Oller, J. A.",
    title = "{Clues for the existence of two K(1)(1270) resonances}",
    eprint = "hep-ph/0610217",
    archivePrefix = "arXiv",
    doi = "10.1103/PhysRevD.75.014017",
    journal = "Phys. Rev. D",
    volume = "75",
    pages = "014017",
    year = "2007"
}

@article{Kolomeitsev:2003ac,
    author = "Kolomeitsev, E. E. and Lutz, M. F. M.",
    title = "{On Heavy light meson resonances and chiral symmetry}",
    eprint = "hep-ph/0307133",
    archivePrefix = "arXiv",
    reportNumber = "GSI-PREPRINT-2003-20",
    doi = "10.1016/j.physletb.2003.10.118",
    journal = "Phys. Lett. B",
    volume = "582",
    pages = "39--48",
    year = "2004"
}

@article{Guo:2006fu,
    author = "Guo, Feng-Kun and Shen, Peng-Nian and Chiang, Huan-Ching and Ping, Rong-Gang and Zou, Bing-Song",
    title = "{Dynamically generated $0^+$ heavy mesons in a heavy chiral unitary approach}",
    eprint = "hep-ph/0603072",
    archivePrefix = "arXiv",
    doi = "10.1016/j.physletb.2006.08.064",
    journal = "Phys. Lett. B",
    volume = "641",
    pages = "278--285",
    year = "2006"
}

@article{Guo:2006rp,
    author = "Guo, Feng-Kun and Shen, Peng-Nian and Chiang, Huan-Ching",
    title = "{Dynamically generated $1^+$ heavy mesons}",
    eprint = "hep-ph/0610008",
    archivePrefix = "arXiv",
    doi = "10.1016/j.physletb.2007.01.050",
    journal = "Phys. Lett. B",
    volume = "647",
    pages = "133--139",
    year = "2007"
}

@article{Molina:2023uko,
    author = "Molina, R. and Liang, Wei-Hong and Xiao, Chu-Wen and Sun, Zhi-Feng and Oset, E.",
    title = "{Two states for the $\Xi(1820)$ resonance}",
    eprint = "2309.03618",
    archivePrefix = "arXiv",
    primaryClass = "hep-ph",
    month = "9",
    year = "2023",
    journal = "arXiv preprint"
}

@article{Hyodo:2011ur,
    author = "Hyodo, Tetsuo and Jido, Daisuke",
    title = "{The nature of the Lambda(1405) resonance in chiral dynamics}",
    eprint = "1104.4474",
    archivePrefix = "arXiv",
    primaryClass = "nucl-th",
    doi = "10.1016/j.ppnp.2011.07.002",
    journal = "Prog. Part. Nucl. Phys.",
    volume = "67",
    pages = "55--98",
    year = "2012"
}

@article{Zhuang:2026lta,
    author = "Zhuang, Zejian and Dom{\'\i}nguez, Fernando Gil and Molina, Raquel",
    title = "{Revealing the $D_0^*(2300)$ two-pole structure from lattice data and the SU(3) limit}",
    eprint = "2606.03826",
    archivePrefix = "arXiv",
    primaryClass = "hep-ph",
    month = "6",
    year = "2026",
    journal = "arXiv preprint"
}

@article{Godfrey:1985xj,
    author = "Godfrey, Stephen and Isgur, Nathan",
    title = "{Mesons in a Relativized Quark Model with Chromodynamics}",
    doi = "10.1103/PhysRevD.32.189",
    journal = "Phys. Rev. D",
    volume = "32",
    pages = "189--231",
    year = "1985"
}

@article{DiPierro:2001uu,
    author = "Di Pierro, Massimo and Eichten, Estia",
    title = "{Excited Heavy-Light Systems and Hadronic Transitions}",
    eprint = "hep-ph/0104208",
    archivePrefix = "arXiv",
    doi = "10.1103/PhysRevD.64.114004",
    journal = "Phys. Rev. D",
    volume = "64",
    pages = "114004",
    year = "2001"
}

@article{Godfrey:2003kg,
    author = "Godfrey, Stephen",
    title = "{Testing the Nature of the $D_{sJ}^*(2317)^+$ and $D_{sJ}(2463)^+$ States Using Radiative Transitions}",
    eprint = "hep-ph/0305122",
    archivePrefix = "arXiv",
    doi = "10.1016/j.physletb.2003.06.049",
    journal = "Phys. Lett. B",
    volume = "568",
    pages = "254--260",
    year = "2003"
}

@article{Godfrey:2005wwa,
    author = "Godfrey, Stephen",
    title = "{Properties of the Charmed P-wave Mesons}",
    eprint = "hep-ph/0508078",
    archivePrefix = "arXiv",
    doi = "10.1103/PhysRevD.72.054029",
    journal = "Phys. Rev. D",
    volume = "72",
    pages = "054029",
    year = "2005"
}

@article{Cleven:2014oka,
    author = "Cleven, Martin and Griesshammer, Harald W. and Guo, Feng-Kun and Hanhart, Christoph and Meissner, Ulf-G.",
    title = "{Strong and Radiative Decays of the $D_{s0}^*(2317)$ and $D_{s1}(2460)$}",
    eprint = "1405.2242",
    archivePrefix = "arXiv",
    primaryClass = "hep-ph",
    doi = "10.1140/epja/i2014-14149-y",
    journal = "Eur. Phys. J. A",
    volume = "50",
    pages = "149",
    year = "2014"
}

@article{Yang:2021tvc,
    author = "Yang, Zhi and Wang, Guang-Juan and Wu, Jia-Jun and Oka, Makoto and Zhu, Shi-Lin",
    title = "{Novel Coupled Channel Framework Connecting the Quark Model and Lattice QCD: An Investigation on Near-Threshold $D_s$ States}",
    eprint = "2107.04860",
    archivePrefix = "arXiv",
    primaryClass = "hep-ph",
    doi = "10.1103/PhysRevLett.128.112001",
    journal = "Phys. Rev. Lett.",
    volume = "128",
    pages = "112001",
    year = "2022"
}

@article{Mohler:2012na,
    author = "Mohler, Daniel and Prelovsek, Sasa and Woloshyn, R. M.",
    title = "{$D\pi$ Scattering and $D$ Meson Resonances from Lattice QCD}",
    eprint = "1208.4059",
    archivePrefix = "arXiv",
    primaryClass = "hep-lat",
    doi = "10.1103/PhysRevD.87.034501",
    journal = "Phys. Rev. D",
    volume = "87",
    pages = "034501",
    year = "2013"
}

@article{Gayer:2021xzv,
    author = "Gayer, Luke and Lang, Nicolas and Ryan, Sinead M. and Tims, David and Thomas, Christopher E. and Wilson, David J.",
    title = "{Isospin-$1/2$ $D\pi$ Scattering and the Lightest $D_0^*$ Resonance from Lattice QCD}",
    eprint = "2102.04973",
    archivePrefix = "arXiv",
    primaryClass = "hep-lat",
    doi = "10.1007/JHEP07(2021)123",
    journal = "JHEP",
    volume = "07",
    pages = "123",
    year = "2021"
}

@article{Yan:2024gkx,
    author = "Yan, Haobo and Liu, Chuan and Liu, Liuming and Meng, Yu and Xing, Hanyang",
    title = "{Pion Mass Dependence in $D\pi$ Scattering and the $D_0^*(2300)$ Resonance from Lattice QCD}",
    eprint = "2404.13479",
    archivePrefix = "arXiv",
    primaryClass = "hep-lat",
    doi = "10.1103/PhysRevD.111.014503",
    journal = "Phys. Rev. D",
    volume = "111",
    pages = "014503",
    year = "2025"
}

@article{Luo:2026qen,
    author = "Luo, Jing and Wu, Bing and Shi, Pan-Pan and Du, Meng-Lin",
    title = "{Analysis of the $D_0^*(2300)$ Resonance from Lattice QCD under Chiral Symmetry}",
    eprint = "2604.18286",
    archivePrefix = "arXiv",
    primaryClass = "hep-ph",
    month = "4",
    year = "2026",
    journal = "arXiv preprint"
}

@article{TorresRincon:2023sdf,
    author = "Torres-Rincon, Juan M. and Ramos, Angels and Tolos, Laura",
    title = "{Femtoscopy of $D$ Mesons and Light Mesons upon Unitarized Effective Field Theories}",
    eprint = "2307.02102",
    archivePrefix = "arXiv",
    primaryClass = "hep-ph",
    doi = "10.1103/PhysRevD.108.096008",
    journal = "Phys. Rev. D",
    volume = "108",
    pages = "096008",
    year = "2023"
}

@article{Guo:2009ct,
    author = "Guo, Feng-Kun and Hanhart, Christoph and Meissner, Ulf-G.",
    title = "{Interactions between Heavy Mesons and Goldstone Bosons from Chiral Dynamics}",
    eprint = "0901.1597",
    archivePrefix = "arXiv",
    primaryClass = "hep-ph",
    doi = "10.1140/epja/i2009-10762-1",
    journal = "Eur. Phys. J. A",
    volume = "40",
    pages = "171--179",
    year = "2009"
}

@article{Altenbuchinger:2013vwa,
    author = "Altenbuchinger, Martin and Geng, Li-Sheng and Weise, Wolfram",
    title = "{Scattering Lengths of Nambu--Goldstone Bosons off $D$ Mesons and Dynamically Generated Heavy-Light Mesons}",
    eprint = "1309.4743",
    archivePrefix = "arXiv",
    primaryClass = "hep-ph",
    doi = "10.1103/PhysRevD.89.014026",
    journal = "Phys. Rev. D",
    volume = "89",
    pages = "014026",
    year = "2014"
}

@article{Yao:2015qia,
    author = "Yao, De-Liang and Du, Meng-Lin and Guo, Feng-Kun and Meissner, Ulf-G.",
    title = "{One-Loop Analysis of the Interactions between Charmed Mesons and Goldstone Bosons}",
    eprint = "1502.05981",
    archivePrefix = "arXiv",
    primaryClass = "hep-ph",
    doi = "10.1007/JHEP11(2015)058",
    journal = "JHEP",
    volume = "11",
    pages = "058",
    year = "2015"
}

@article{Du:2016puz,
    author = "Du, Meng-Lin and Guo, Feng-Kun and Meissner, Ulf-G. and Yao, De-Liang",
    title = "{Aspects of the Low-Energy Constants in the Chiral Lagrangian for Charmed Mesons}",
    eprint = "1610.02963",
    archivePrefix = "arXiv",
    primaryClass = "hep-ph",
    doi = "10.1103/PhysRevD.94.094037",
    journal = "Phys. Rev. D",
    volume = "94",
    pages = "094037",
    year = "2016"
}

@article{Lutz:2022dcl,
    author = "Lutz, Matthias F. M. and Guo, Xiao-Yu and Heo, Yonggoo and Korpa, C. L.",
    title = "{Coupled-Channel Dynamics with Chiral Long-Range Forces in the Open-Charm Sector of QCD}",
    eprint = "2209.10601",
    archivePrefix = "arXiv",
    primaryClass = "hep-ph",
    month = "9",
    year = "2022",
    journal = "arXiv preprint"
}

@article{Meissner:2020khl,
    author = "Meissner, Ulf-G.",
    title = "{Two-Pole Structures in QCD: Facts, Not Fantasy!}",
    eprint = "2005.06909",
    archivePrefix = "arXiv",
    primaryClass = "hep-ph",
    doi = "10.3390/sym12060981",
    journal = "Symmetry",
    volume = "12",
    pages = "981",
    year = "2020"
}

@article{Mai:2022eur,
    author = "Mai, Maxim and Meissner, Ulf-G. and Urbach, Carsten",
    title = "{Towards a Theory of Hadron Resonances}",
    eprint = "2206.01477",
    archivePrefix = "arXiv",
    primaryClass = "hep-ph",
    doi = "10.1016/j.physrep.2022.11.005",
    journal = "Phys. Rept.",
    volume = "1001",
    pages = "1--66",
    year = "2023"
}

\end{document}